\documentclass[aps,prb,twocolumn,amsmath,amssymb,floatfix,superscriptaddress]{revtex4-1}
\usepackage{graphicx}
\usepackage{bm}
\usepackage[hidelinks]{hyperref}

\usepackage{braket,bbold,color,textcomp,gensymb}

\newcommand{\bs}[1]{\boldsymbol{#1}}
\begin{document}

\title{Trions in Twisted Bilayer Graphene}

\author{Frank Schindler}
\affiliation{Princeton Center for Theoretical Science, Princeton University, Princeton, NJ 08544, USA}

\author{Oskar Vafek}
\affiliation{National High Magnetic Field Laboratory, Tallahassee, Florida, 32310, USA}
\affiliation{Department of Physics, Florida State University, Tallahassee, Florida 32306, USA}

\author{B. Andrei Bernevig}
\affiliation{Department of Physics, Princeton University, Princeton, NJ 08544, USA}
\affiliation{Donostia International Physics Center, P. Manuel de Lardizabal 4, 20018 Donostia-San Sebastian, Spain}
\affiliation{IKERBASQUE, Basque Foundation for Science, Bilbao, Spain}

\begin{abstract}
The strong coupling phase diagram of magic angle twisted bilayer graphene (TBG) predicts a series of  exact one particle charge $\pm 1$ gapped  excitations on top of the integer-filled ferromagnetic ground-states. Finite-size exact diagonalization studies showed that these are {\it{the lowest} } charge $\pm 1$ excitations in the system (for $10\text{nm}$ screening length), with the exception of charge $+1$ at filling $-1$ in the chiral limit. In the current paper we show that this ``trion bound state", a $3$-particle, charge $1$ excitation of the insulating ferromagnetic ground-state of the projected Hamiltonian of TBG is the lowest charge $+1$ overall excitation at  $\nu=-1$, and also for some large ($\approx 20\text{nm}$) screening lengths at $\nu=-2$ in the chiral limit and with very small binding energy. At other fillings, we show that trion bound states do exist, but only for momentum  ranges that do not cover the entire moiré Brillouin zone. The trion bound states (at different momenta) exist for finite parameter range $w_0/w_1$ but they all disappear in the continuum far below the realistic values of $w_0/w_1= 0.8$. We find the conditions for the existence of the trion bound state, a good variational wavefunction for it, and investigate its  behavior for different screening lengths, at all integer fillings, on both the electron and hole sides. 
\end{abstract}
\maketitle

\section{Introduction}
Cascades of transitions observed at integer fillings of the narrow bands of the magic angle twisted bilayer graphene (TBG) in scanning tunneling spectroscopy (STM)~\cite{choi_imaging_2019,kerelsky_2019_stm,choi2020tracing,xie2019spectroscopic,Wong2020Nature,choi2021interactiondriven} as well as in electronic compressibility~\cite{Shahal2020Nature,burg2020evidence,saito2020isospin,Yacoby2021NatPhys,BenFeldman2021} experiments have demonstrated the strongly correlated nature of this remarkable physical system~\cite{cao_correlated_2018,cao_strange_2020,cao_tunable_2020,cao_unconventional_2018,lu2019superconductors,liu_spin-polarized_2019,pixley2019,polshyn_linear_2019,yankowitz2019tuning,das2020symmetry,park2020flavour,nuckolls_chern_2020,wu_chern_2020,rozen2020entropic,stepanov_interplay_2020,cao2021,Yacoby2021,Kim21,MacD2014Abinitio,Uchida_corrugation,jain_corrugation,kang_symmetry_2018,zou2018,fu2018magicangle,yuan2018model,tarnopolsky_origin_2019,hejazi_multiple_2019,serlin_QAH_2019,po_faithful_2019,zhang2019nearly,song_all_2019,liu2019quantum,Wilson2020TBG,koshino2020_effective,hejazi2020hybrid,Katz2020,wang2020chiral,lu2020fingerprints,Kwan2020ParallelMagnetic,padhi2020transport,kwan2020excitonic,davydov2020four,Kwan2021Excitons,Kwan2021DomainWalls,kwan2021kekule,wagner2021global,Portol_s_2021,wang2021onedimensional,Rickhaus2021,Libisch21,kim2021spectroscopic,chaudhary2021shiftcurrent,ledwith2021tb,ledwith2021family,li2021magicangle}.
The spectroscopic shifts observed at integer fillings in the narrow band's tunneling density of states observed in STM are comparable to, or larger than, the narrow bandwidth. Moreover, there are clear spectral signatures of a high density of states approaching the Fermi level as a non-zero integer filling $\nu$ is reached from the charge neutrality point (CNP) side, but then a rapid resetting to $\sim 20$meV above the Fermi level is observed as the same $\nu$ is approached from the remote band side~\cite{Wong2020Nature}. This phenomenology is naturally understood as a cascade between light and heavy fermions~\cite{KangBernevigVafek2021}, where the electron (hole) excitations at a positive (negative) integer $\nu$ are light fermions with a minimum at the center of the moiré Brillouin zone (mBZ), while the hole (electron) excitations at a positive (negative)  integer $\nu$ are heavy fermions~\cite{TBG5,VafekKangPRB2021}.

In this paper, we extend the analysis of the strong coupling limit of TBG~\cite{koshino_maximally_2018,wux2018b,venderbos2018,thomson2018triangular,ochi_possible_2018,xux2018,lee2018emergent,dodaro2018phases,sharpe_emergent_2019,liu2019pseudo,jiang_charge_2019,burg_correlated_2019,xie_HF_2020,wu_collective_2020,liu2020tuning,daliao_VBO_2019,classen2019competing,huang2019antiferromagnetically,seo_ferro_2019,repellin_EDDMRG_2020,repellin_FCI_2020,Xie2020TBG,ledwith2020,zhang_HF_2020,abouelkomsan2020,cea_band_2020,soejima2020efficient,VafekKangPRL2020,liu2020nematic,liu2020theories,daliao2020correlation,chen_tunable_2020,chichinadze2020prb,khalaf2020soft,fernandes_nematic_2020,kang_nonabelian_2020,saito_independent_2020,saito2020,Chichinadze2020Nematic,thomson2021gatedefined,Thomson21,Kumar21,chichinadze2021su4,Samajdar_2021,hofmann2021fermionic,StrongCouplingLectureNotes} at integer filling by studying a further set of excitations - trions, \emph{i.e.}, {\it composite} charge $\pm 1$ excitations, consisting of two electrons and one hole, or two holes and one electron~\cite{zeng2021strong,turner2021gapping}. We specifically look for bound states below the $2$ electron -- $1$ hole (or vice-versa) continuum. On the light mass side, we find that trion bound states are at a higher energy than the lowest single particle excitations for a range of momenta near the minimum of the dispersion, but can become lower than single particle excitations for larger momenta and higher energies. 
On the heavy mass side, the trion bound state energy can be very close to the lowest energy single particle charge excitation for a range of momenta. Although we never find the trion bound state to be the absolute lowest energy excitation for realistic values of TBG parameters at any integer $\nu$, in the idealized chiral limit trion bound states are lower in energy at all momenta with a small binding energy on the heavy mass side of $\nu=\pm 1$, where single particle excitations form an almost perfectly flat band (Fig. 6\textbf{b} of Ref.~\onlinecite{TBG5})~\cite{KangBernevigVafek2021,VafekKangPRB2021}. 
The absence of a lower energy excitation than the single particle excitation (recently also pointed out in Ref.~\onlinecite{kwan2021skyrmions} for realistic TBG parameters) for $|\nu|=2$ (for screening length $\xi=10\text{nm}$ at all $w_0/w_1=0$) has negative implications for associating this excitation with the superconducting mechanism~\cite{po_origin_2018,Wu2018TBG-BCS,isobe2018unconventional,guo2018pairing,xu2018topological,guinea2018,kennes2018strong,liu2018chiral,wu_phonon_linearT2019,Lian2019TBG,gonzalez2019kohn,you2019,hu2019_superfluid,xie_superfluid_2020,julku_superfluid_2020,arora_2020,knig2020spin,khalaf2020symmetry,lewandowski2020pairing,Stepanov21,huang2021pseudospin,Fernandes21,Fernandes21PRB,you2021kohnluttinger}. In particular, the existence of trion bound states is a necessary requirement for the formation of skyrmions, which have been hypothesized as a pairing glue for TBG~\cite{christos2020superconductivity,Khalaf21,kwan2021skyrmions}. While skyrmions may be stabilized in Chern bands~\cite{PhysRevB.47.16419,PhysRevB.50.11018,PhysRevB.56.6795,PhysRevB.54.R2296} -- such as the flat bands of TBG in the chiral limit, $w_0/w_1=0$ -- the absence of trion bound states at realistic parameter values implies that they do not persist away from the chiral limit (see also \emph{Note added}).

Our results are consistent with the previous exact diagonalization (ED) study~\cite{TBG6} with $\xi=10$nm in which single-particle excitations were found to have the lowest absolute energy, except in the chiral limit at $|\nu|=1$ and filling towards the CNP, \emph{i.e.}, on the heavy mass side. In Fig.~14\textbf{b} of Ref.~\onlinecite{TBG6}, a trion bound state composed of 2 electrons and 1 hole, all in the same Chern sector, is shown to have a lower energy than the single (heavy) electron excitation at $\nu=-1$ by about $1 \mathrm{meV}$. At $\nu=-1$, but for opposite charge -- and in both charge sectors of all the other integer fillings -- the ED study~\cite{TBG6} shows that the one-particle excitation is the lowest excitation of the system. Ref.~\onlinecite{TBG6} did not, however, investigate the existence of trion bound states at specific momenta or away from the chiral limit, nor did it obtain analytic variational solutions for these states, which we now perform here. 
By performing momentum-resolved calculations, here we show that at $\nu=0, - 1,-2,-3 $ and in the charge $-1$ sector,  there is no 2 hole -- 1 electron trion bound state at, or close to, ${\bf k}=0$ (i.e the light 1-particle minimum) below the 2 hole -- 1 electron continuum, but a bound state develops at \emph{finite} momenta and higher energy for $w_0/w_1<0.6$ (see Fig.~\ref{figmain: projectedspectra}\textbf{b},\textbf{d},\textbf{f} below); it disappears for larger $w_0/w_1$. Here charge $+1$ is taken to be the charge of the electron. At $\nu= 0,-2$ (for $\xi=10\text{nm}$),$-3$ and in the charge $+1$ sector, there is no trion bound state at, or close to, ${\bf k}=K,M$ (i.e. the heavy 1-particle minimum)  below the 2 electron -- 1 hole continuum, but a bound state at \emph{finite} momenta away from $K,M$ develops for $w_0/w_1<0.6$ (see Fig.~\ref{figmain: exactspectra} and Fig.~\ref{figmain: exactchiralspectra}\textbf{b} below), disappearing for larger $w_0/w_1$. In the chiral limit, the $\nu=-1$ charge $+1$ sector is the only one exhibiting a clear trion bound state with a significant gap over the entire BZ (see Fig.~\ref{figmain: exactchiralspectra}\textbf{a} below), in agreement with Ref.~\onlinecite{TBG6}. 

We provide analytic arguments for the non-existence of a full trion bound state when the quasiparticle mass is significantly lighter than that of the Goldstone, which we test by artificially tuning  the $\nu=-1$, charge $+1$ excitation lower and seeing the bound state disappear. We also provide a variational wavefunction (with large overlap) of the trion bound states.

\begin{figure}[t]
\includegraphics[width=0.48\textwidth]{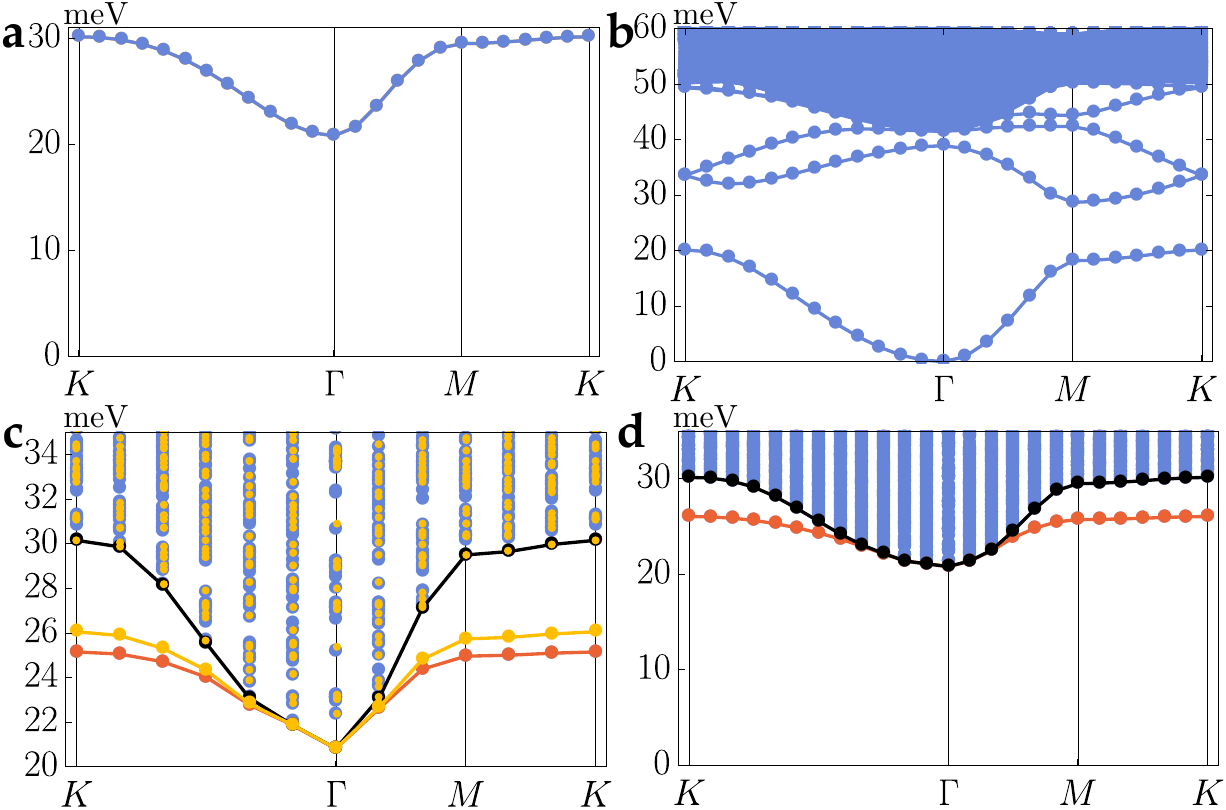}
\caption{Quasiparticle-Goldstone excitations and projected trion Hamiltonian in the chiral limit and at $\nu=0$. \textbf{a}~Single-particle ($N=1, Q=1$) energy. \textbf{b}~Goldstone mode ($N=2, Q=0$) energy. \textbf{c}~Comparison of exact (blue, bound state in red) and quasiparticle-Goldstone projected (yellow) $N=3,Q=1$ spectra (Fig.~\ref{figmain: exactspectra}\textbf{a}, zoomed-in) for a fixed mBZ. \textbf{d}~Quasiparticle-Goldstone projected $N=3,Q=1$ spectrum for a finer mBZ.}
\label{figmain: goldstoneprojection}
\end{figure}

\section{Trion eigenvalue problem in TBG}
When projected into magic angle flat bands, the interacting Hamiltonian of TBG becomes a positive-semidefinite Hamiltonian~\cite{KangVafek19, TBG3, Bultinck2020PRX}:
\begin{equation} \label{eqmain: nonchiral-flat Hamiltonian}
    H = \frac{1}{2\Omega} \sum_{\bs{q},\bs{G}} O^\dagger_{\bs{q}, \bs{G}} O_{\bs{q},\bs{G}},
\end{equation}
\begin{equation}
\begin{aligned}
O_{\bs{q}, \bs{G}} =& \sqrt{V(\bs{q}+\bs{G})} \sum_{\bs{k},m,n,\eta,s} M^{(\eta)}_{m,n} (\bs{k},\bs{q}+\bs{G}) \\&\left(c^\dagger_{\bs{k} + \bs{q},m,\eta,s} c_{\bs{k},n,\eta,s} - \frac{1}{2} \delta_{\bs{q},\bs{0}} \delta_{m,n} \right).
\end{aligned}
\end{equation}
Here, $\bs{q}$ ranges over the mBZ, and $\bs{G}$ ranges over the moir\'e reciprocal lattice $\mathcal{Q}_0$ defined in Ref.~\onlinecite{TBG1}.  $c^\dagger_{\bs{k},n,\eta,s}$ creates an electron at moir\'e momentum $\bs{k}$ in the eigenstate $n=\pm$ of the Bistritzer-MacDonald (BM) continuum Hamiltonian~\cite{Bistritzer11,Efimkin2018TBG} for graphene valley $\eta = \pm$ and spin $s = \pm$. $V(\bs{q})$ is the double gate-screened Coulomb interaction~\cite{TBG3}. The form factors are $ \label{eqmain: formfactors}
M^{(\eta)}_{m,n} (\bs{k},\bs{q}+\bs{G}) = \sum_{\alpha, \bs{Q}} u^*_{\bs{Q}-\bs{G},\alpha;m,\eta}(\bs{k} + \bs{q}) u_{\bs{Q},\alpha;n,\eta}(\bs{k})$,
where $\alpha$ ranges over the two graphene sublattices $A$ and $B$, $\bs{Q}$ ranges over the hexagonal momentum lattice $\mathcal{Q}_\pm$ defined in Ref.~\onlinecite{TBG1}, and $u_{\bs{Q},\alpha;n,\eta}(\bs{k})$ are the BM Bloch states.
Since $H$ is positive-semidefinite, it is possible to find exact ground states, in certain limits at integer filling, which are U(4) ferromagnets~\cite{KangVafek19,alavirad2019ferromagnetism,Bultinck2020PRL,Bultinck2020PRX,TBG4}. Specifically, while these states are always exact ground states at filling $\nu=0$, they become exact ground-states at $\nu=\pm 2$ when the bands satisfy certain quantum geometry, and at $\nu=\pm 1,3$ in the chiral limit $w_0=0$~\cite{TBG1,TBG4}. For simplicity, we first focus on this limit here, and then analyze the far interpolation to the realistic situation. Then, the form factors become diagonal in the Chern basis
$d^\dagger_{\bs{k},e_Y,\eta,s} = {(c^\dagger_{\bs{k},+,\eta,s}+\mathrm{i} e_Y c^\dagger_{\bs{k},-,\eta,s})}/{\sqrt{2}}$
and are independent of the valley-index, so that we denote them by $M^{(e_Y)}(\bs{k},\bs{q}+\bs{G})$.

Since the ground state is a U(4) ferromagnet, the Hamiltonian $H$ preserves not only the charge $Q$ and total momentum $\bs{p}$ of excitations, but also the total number of electron creation/annihilation operators $N$. The $N=3$ trion scattering matrix for charge $Q=+1$ follows from acting with $H$ on the excitation $d^\dagger_{\bs{k}_3,e_Y'',\eta_3,s_3} d^\dagger_{\bs{k}_2,e_Y',\eta_2,s_2} d_{\bs{k}_{23},e_Y,\eta_1,s_1} \ket{\Psi}$, and we define $\bs{k}_{23} = \bs{k}_2+\bs{k}_3-\bs{p}$ as $H$ preserves total momentum:
\begin{equation} \label{eqmain: trionscattmat_chiral}
\begin{aligned}
&H d^\dagger_{\bs{k}_3,e_Y'',\eta_3,s_3} d^\dagger_{\bs{k}_2,e_Y',\eta_2,s_2} d_{\bs{k}_{23},e_Y,\eta_1,s_1} \ket{\Psi}=\\
&\sum_{\tilde{\bs{k}}_3,\tilde{\bs{k}}_2} W^{(e_Y'',e_Y',e_Y)}_{\tilde{\bs{k}}_3,\tilde{\bs{k}}_2; \bs{k}_3,\bs{k}_2} (\bs{p}) d^\dagger_{\tilde{\bs{k}}_3,e_Y'',\eta_3,s_3} d^\dagger_{\tilde{\bs{k}}_2,e_Y',\eta_2,s_2} d_{\tilde{\bs{k}}_{23},e_Y,\eta_1,s_1} \ket{\Psi}.
\end{aligned}
\end{equation}
For $\ket{\Psi} = \ket{\Psi_0}$ the ground state at $\nu=0$~\cite{TBG3}, we find the $N_{\mathcal{M}}^2 \times N_{\mathcal{M}}^2$ scattering matrix
\begin{widetext}
\begin{equation} \label{eqmain: qmatdef_chiral}
\begin{aligned}
W^{(e_Y'',e_Y',e_Y)}_{\tilde{\bs{k}}_3,\tilde{\bs{k}}_2; \bs{k}_3,\bs{k}_2} (\bs{p}) =& \delta_{\bs{k}_3,\tilde{\bs{k}}_3} \delta_{\bs{k}_2,\tilde{\bs{k}}_2} \bigg[R (\bs{k}_3) + R (\bs{k}_2) + R (\bs{k}_2+\bs{k}_3-\bs{p}) \bigg] -2 \delta_{\tilde{\bs{k}}_3,\bs{k}_3} S^{(e_Y',e_Y)}_{\tilde{\bs{k}}_2+\bs{k}_3-\bs{p};\bs{k}_3+\bs{k}_2-\bs{p}} (\bs{p}-\bs{k}_3)
\\&
+ 2 \delta_{\tilde{\bs{k}}_2+\tilde{\bs{k}}_3,\bs{k}_2+\bs{k}_3} S^{(e_Y'',e_Y')*}_{\bs{k}_2+\bs{k}_3-\tilde{\bs{k}}_2;\bs{k}_3} (\tilde{\bs{k}}_2-\bs{k}_3)
-2 \delta_{\tilde{\bs{k}}_2,\bs{k}_2} S^{(e_Y'',e_Y)}_{\tilde{\bs{k}}_3+\bs{k}_2-\bs{p};\bs{k}_3+\bs{k}_2-\bs{p}} (\bs{p}-\bs{k}_2).
\end{aligned}
\end{equation}
\end{widetext}
We have used the chiral limit $N=1,Q=1$ dispersion
\begin{equation} \label{eqmain: rmatdef_chiral}
    R (\bs{k}) = \frac{1}{2\Omega} \sum_{\bs{q},\bs{G}} V(\bs{q}+\bs{G}) |M^{(e_Y)}(\bs{k},\bs{q}+\bs{G})|^2, 
\end{equation}
and $N=2,Q=0$ scattering matrix~\cite{VafekKangPRB2021,TBG5}
\begin{equation}
\begin{aligned}
S^{(e_Y,e_Y')}_{\tilde{\bs{k}};\bs{k}} (\bs{p}) =& \frac{1}{2\Omega} \sum_{\bs{G}} V(\bs{k}-\tilde{\bs{k}}+\bs{G}) \\M^{(e_Y)*}&(\tilde{\bs{k}}+\bs{p},\bs{k}-\tilde{\bs{k}}+\bs{G}) M^{(e_Y')} (\tilde{\bs{k}},\bs{k}-\tilde{\bs{k}}+\bs{G}).
\end{aligned}
\end{equation}
The appearance of these terms in the $N=3,Q=1$ trion scattering matrix can be understood as follows: at charge neutrality, electrons and holes follow the same dispersion relation, and so both the two electrons and the hole each contribute one $R$ matrix term to $W$ that captures their kinetic energy. Moreover, each of the two electrons may interact with the hole via the Coulomb attraction that is encapsulated in the two $S$ matrix terms. Finally, the two electrons may interact with each other via the Coulomb repulsion effected by the $S^*$ matrix term.
Similar expressions can be derived for general ground states $\ket{\Psi}$~\cite{TBG5}. Numerically, we find the lowest trion bound states in the Hilbert space sectors where all chiralities are equal in Eq.~\eqref{eqmain: trionscattmat_chiral}: $e_Y''=e_Y'=e_Y$. Furthermore, we must choose the spin-valley flavors $(\eta_3,s_3)$ and $(\eta_2,s_2)$ so that they are unoccupied in $\ket{\Psi}$, and $(\eta_1,s_1)$ so that it is occupied in $\ket{\Psi}$. We have checked that the choice $(\eta_3,s_3)=(\eta_2,s_2)$ yields the same bound states as $(\eta_3,s_3)\neq(\eta_2,s_2)$, which we focus on in the following.

\begin{figure}[t]
\includegraphics[width=0.48\textwidth]{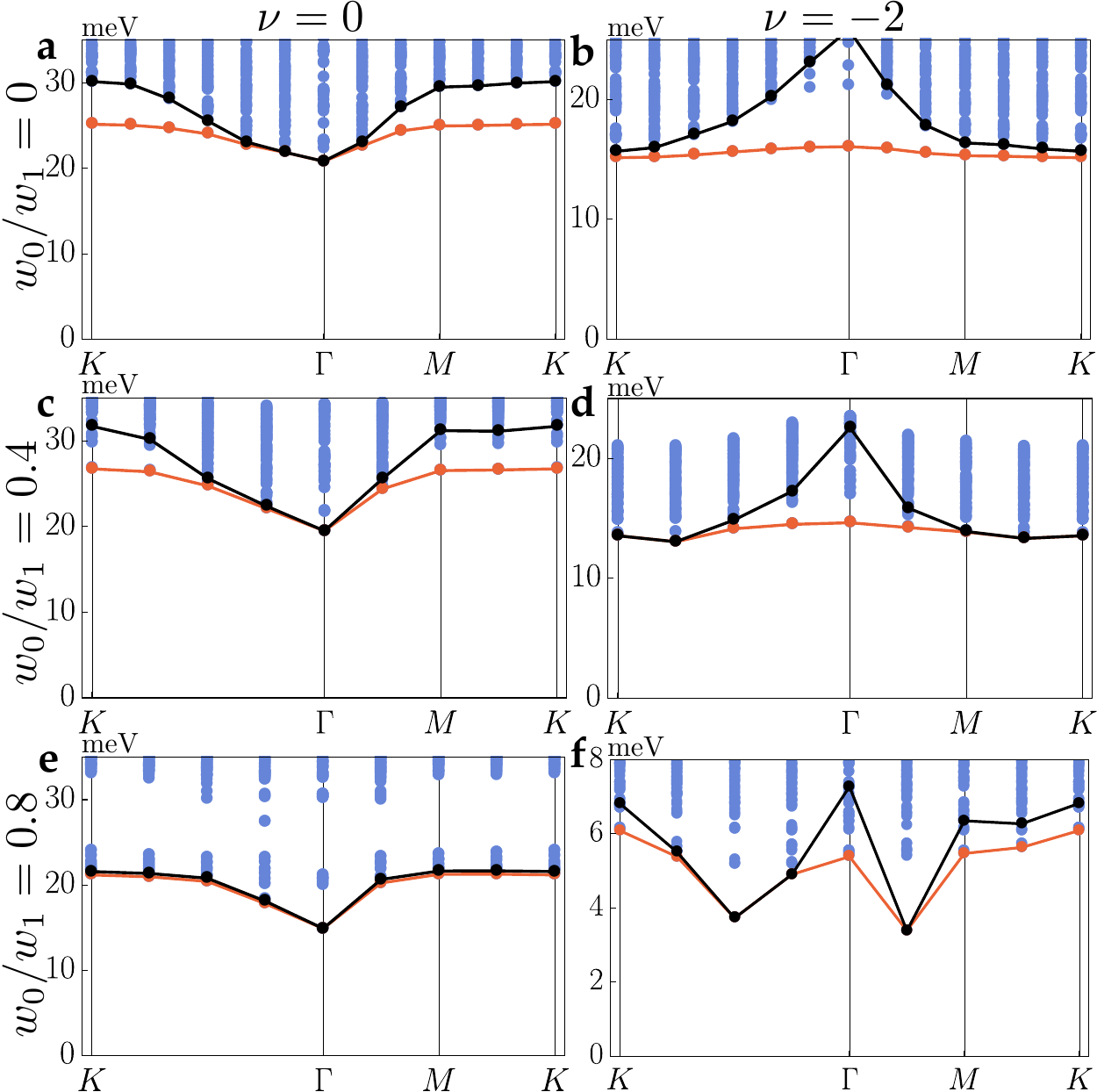}
\caption{Exact $N=3, Q=1$ spectra with minimum of quasiparticle-Goldstone continuum highlighted (black) for fillings $\nu=0,-2$ and different values of $w_0/w_1$ (only the lowest 100 states are shown). For $\nu=0$, where $Q=1$ implies doping away from charge neutrality, a trion bound state develops away (but not at) from $\Gamma$ close to the chiral limit ($w_0/w_1<0.6$). For $\nu=-2$, where $Q=1$ implies doping towards charge neutrality, trion bound state develops at $\Gamma$. It has weak binding energy at the band edge only extremely close to the chiral limit.
We note that exact ground states at fillings $\nu=-1,-3$ are not known for finite values of $w_0/w_1$~\cite{TBG4}, however, the corresponding exact spectra in the chiral limit $w_0/w_1 = 0$ can be computed and are shown in Fig.~\ref{figmain: exactchiralspectra}.
Our model parameters are defined in Ref.~\onlinecite{TBG5} and given by $\theta = 1.05\degree$, $v_F = 5.944 \mathrm{eV}$\AA, $|K| = 1.703$\AA$^{-1}$, $w_1 = 110 \mathrm{meV}$, $U_\xi = 26 \mathrm{meV}$, and $\xi = 20 \mathrm{nm}$.
}
\label{figmain: exactspectra}
\end{figure}

\begin{figure}[t]
\includegraphics[width=0.48\textwidth]{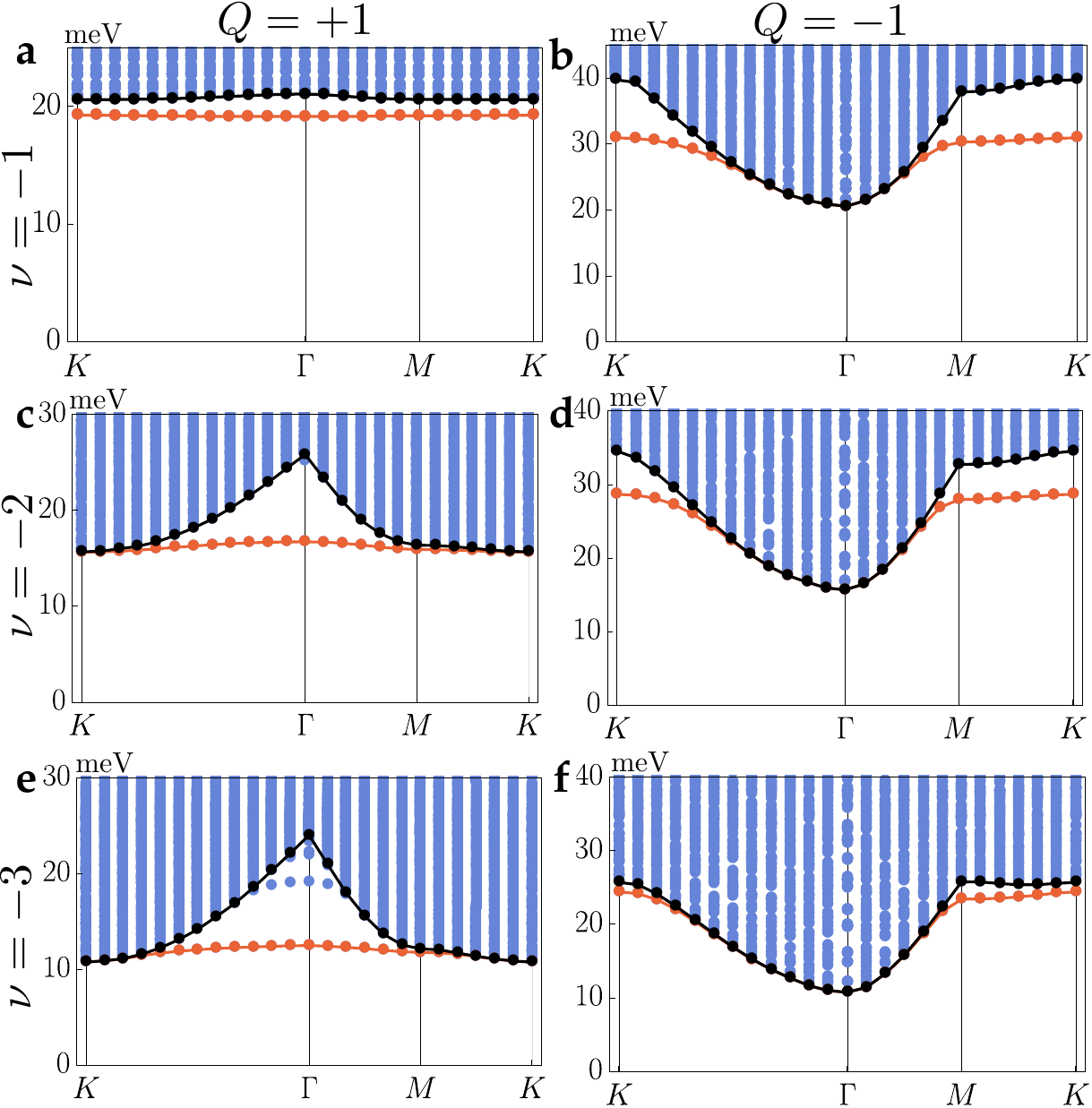}
\caption{Chiral limit trion spectra obtained after projecting the $N=3$ scattering matrix into low-energy quasiparticle-Goldstone states, with minimum of quasiparticle-Goldstone continuum highlighted (black). See Fig.~\ref{figmain: exactspectra} for model parameters. For $\nu=-1, Q=1$ (panel \textbf{a}), the trion bound state persists at all momenta (but due to its small binding energy likely disappears outside the chiral limit).}
\label{figmain: projectedspectra}
\end{figure}

\section{Quasiparticle-Goldstone approximation}
The low energy part of the $N=3$, $Q=1$ Hilbert space is dominated by the combination of charge $+1$ ($N=1$, $Q=1$) quasiparticles and Goldstone ($N=2$, $Q=0$) excitations, which are plotted in Figs.~\ref{figmain: goldstoneprojection}\textbf{a},\textbf{b}, respectively, for $\nu=0$. In Fig.~\ref{figmain: exactspectra}, these form the quasiparticle-Goldstone continuum (shown above the black line). The quasiparticle-Goldstone product states read
\begin{equation} \label{eqmain: varbasis}
\begin{aligned}
\ket{\Phi^{(e_Y',e_Y)}_{\bs{p};\bs{q},1}} = \sum_{\bs{k}} G^{(e_Y)}_{\bs{p}-\bs{q},\bs{k}} d^\dagger_{\bs{q},e_Y',\eta_3,s_3} &d^\dagger_{\bs{k},e_Y,\eta_2,s_2}\\ &d_{\bs{q}+\bs{k}-\bs{p},e_Y,\eta_1,s_1} \ket{\Psi}, \\
\ket{\Phi^{(e_Y',e_Y)}_{\bs{p};\bs{q},2}} = \sum_{\bs{k}} G^{(e_Y)}_{\bs{p}-\bs{q},\bs{k}} d^\dagger_{\bs{k},e_Y',\eta_3,s_3} &d^\dagger_{\bs{q},e_Y,\eta_2,s_2}\\ &d_{\bs{q}+\bs{k}-\bs{p},e_Y,\eta_1,s_1} \ket{\Psi},
\end{aligned}
\end{equation}
where $G^{(e_Y)}_{\bs{p}-\bs{q},\bs{k}}$ is the Goldstone wavefunction at total momentum $\bs{p}-\bs{q}$~\cite{TBG5}. While $\ket{\Phi^{(e_Y',e_Y)}_{\bs{p};\bs{q},1}}$ diagonalizes the first line of the trion scattering matrix in Eq.~\eqref{eqmain: qmatdef_chiral}, and $\ket{\Phi^{(e_Y',e_Y)}_{\bs{p};\bs{q},2}}$ diagonalizes the first and last term of Eq.~\eqref{eqmain: qmatdef_chiral}, both states are mixed by the respectively remaining terms. Nevertheless, since the combination of charge $-1$ ($N_{\mathrm{tot}}=1$, $Q_{\mathrm{tot}}=-1$) and charge $+2$ ($N_{\mathrm{tot}}=2$, $Q_{\mathrm{tot}}=2$) excitations is relatively costly, and the lowest Goldstone mode is fully gapped from all higher (including neutral bound state) excitations (Fig.~\ref{figmain: goldstoneprojection}\textbf{b}), it is a good low-energy approximation to project the $N_{\mathrm{tot}}=3$, $Q_{\mathrm{tot}}=1$ scattering matrix into the variational particle-Goldstone basis of Eq.~\eqref{eqmain: varbasis}. Here, we must take into account that this basis is not orthogonal (although it is complete), so that the projected trion Hamiltonian reads
\begin{equation} \label{eqmain: chiral_limit_trions}
\begin{aligned}
\tilde{\mathcal{H}}^{(\bs{p},e_Y',e_Y)} &= \left[\mathcal{O}^{(\bs{p},e_Y',e_Y)}\right]^{-1} \mathcal{H}^{(\bs{p},e_Y',e_Y)}, \\
\mathcal{O}^{(\bs{p},e_Y',e_Y)}_{\bs{q},\alpha;\bs{q'},\beta} &= \braket{\Phi^{(e_Y',e_Y)}_{\bs{p};\bs{q},\alpha}| \Phi^{(e_Y',e_Y)}_{\bs{p};\bs{q},\beta}}, \\
\mathcal{H}^{(\bs{p},e_Y',e_Y)}_{\bs{q},\alpha;\bs{q'},\beta} &= \braket{\Phi^{(e_Y',e_Y)}_{\bs{p};\bs{q},\alpha}|H|\Phi^{(e_Y',e_Y)}_{\bs{p};\bs{q},\beta}},
\end{aligned}
\end{equation}
where $\alpha=1,2$ runs over the two sets of variational states in Eq.~\eqref{eqmain: varbasis}. This matrix is not Hermitian but its spectrum is real and coincides with the variational energies. Moreover, its right eigenvectors are the variational states in the non-orthogonal basis of Eq.~\eqref{eqmain: varbasis}. We have decided to use the orthonormalization in Eq.~\eqref{eqmain: chiral_limit_trions}, rather than the more standard orthonormalization $$\left[\mathcal{O}^{(\bs{p},e_Y',e_Y)}\right]^{-1/2} \mathcal{H}^{(\bs{p},e_Y',e_Y)} \left[\mathcal{O}^{(\bs{p},e_Y',e_Y)}\right]^{-1/2},$$ because the eigenstates of the latter do not correspond to coefficients in the basis of Eq.~\eqref{eqmain: varbasis}. Instead, they correspond to an appropriately orthonormalized basis, which we cannot compute analytically and which is therefore more difficult to interpret. Nevertheless, the physical consequences derived from either orthonormalization prescription are identical. We note that to obtain physical results, all overlaps in the non-orthonormal basis must be computed with the metric $\mathcal{O}^{(\bs{p},e_Y',e_Y)}$ instead of the identity matrix~\cite{PhysRevB.90.075128,PhysRevB.95.115155,doi:10.1063/5.0023185,santos2020nonortho}.

The quasiparticle-Goldstone projection has the numerical advantage that it brings down the computational cost from diagonalizing the $N_{\mathcal{M}}^2 \times N_{\mathcal{M}}^2$ matrix in Eq.~\eqref{eqmain: qmatdef_chiral}, where $N_{\mathcal{M}}$ is the number of momenta in the discretized mBZ, to diagonalizing a $N_{\mathcal{M}} \times N_{\mathcal{M}}$ matrix (the $2N_{\mathcal{M}} \times 2N_{\mathcal{M}}$ matrix of Eq.~\eqref{eqmain: chiral_limit_trions} can be further block-diagonalized into $N_{\mathcal{M}} \times N_{\mathcal{M}}$ blocks, see below). Fig.~\ref{figmain: goldstoneprojection}\textbf{c} shows perfect alignment between the continuum parts of exact and projected spectra, while the exact trion bound state appears slightly below the projected bound state in energy. Hence, this approximation is very good in the continuum, while the projected trion bound state binding energy is slightly decreased.

\begin{table}[t]
\begin{tabular}{|l||l|l|l|l|}
\hline
$(\nu,Q)$                               & (0,1)  & (0,-1)  &  (-1,1) & (-1,-1) \\ \hline \hline
$m^*_{\mathrm{qp}}/m_{\mathrm{e}}$      & 0.098  & 0.098   & 13.06   & 0.048   \\ \hline
$m^*_{\mathrm{G}}/m_{\mathrm{e}}$       & 0.085  & 0.085   & 0.088   & 0.088    \\ \hline
$m^*_{\mathrm{qp}}/m^*_{\mathrm{G}}$    & 1.15   & 1.15    & 148     & 0.54     \\ \hline \hline
$(\nu,Q)$                               & (-2,1) & (-2,-1) & (-3,1)  & (-3,-1) \\ \hline \hline
$m^*_{\mathrm{qp}}/m_{\mathrm{e}}$      & 0.416  & 0.031   & 0.211   & 0.023    \\ \hline
$m^*_{\mathrm{G}}/m_{\mathrm{e}}$       & 0.097  & 0.097   & 0.117   & 0.117   \\ \hline
$m^*_{\mathrm{qp}}/m^*_{\mathrm{G}}$    & 4.28   & 0.32    & 1.80    & 0.20   \\ \hline
\end{tabular}
\caption{\label{tabmain: effectivemasses} Effective masses for quasiparticles ($m^*_{\mathrm{qp}}$) and Goldstone modes ($m^*_{\mathrm{G}}$) in the chiral limit, measured in units of the bare electron mass $m_\mathrm{e}$. See Fig.~\ref{figmain: exactspectra} for model parameters. For doping away from charge neutrality, we calculate $m^*_{\mathrm{qp}}$ from the quasiparticle energy minimum $\Gamma$. For doping towards charge neutrality, we calculate $m^*_{\mathrm{qp}}$ from the quasiparticle energy minimum at $K$.}
\end{table}

\section{Absence of bound state close to the particle-hole minimum for dispersive quasiparticles}
In Figs.~\ref{figmain: exactspectra} and~\ref{figmain: projectedspectra}, the minimum of the trion spectrum is indicated in red, while the minimum of the quasiparticle-Goldstone continuum is indicated in black. Notably, in most cases, the two curves merge at a common minimal energy. Hence, for most choices of $\nu$ and $Q$ -- where there is sizable quasiparticle dispersion larger than that of the Goldstone -- trion bound states only exist away from the absolute minimum of the quasiparticle-Goldstone continuum. The notable exception is the flat quasiparticle case $\nu=-1, Q=1$ (Fig.~\ref{figmain: projectedspectra}\textbf{a}) where a trion bound state persists at all momenta. 

To give an intuition for this, we first note that the states $\ket{\Phi^{(e_Y',e_Y)}_{\bs{p};\bs{p},\alpha}}$ [Eq.~\eqref{eqmain: varbasis}] are \emph{always} eigenstates of $H$ with energy $R(\bs{p})$, because the Goldstone operator at $\Gamma$, $\sum_{\bs{k}} G^{(e_Y)}_{\bs{0},\bs{k}} d^\dagger_{\bs{k},e_Y,\eta,s} d_{\bs{k},e_Y,\eta,s}$, $G^{(e_Y)}_{\bs{0},\bs{k}}= 1/\sqrt{N_\mathcal{M}}$ is a symmetry of the Hamiltonian~\cite{TBG5}. Hence, the momentum at which the quasiparticle-Goldstone continuum assumes its minimum is the same as that of the $Q=1$ dispersion. Without loss of generality, let us here take this minimum to occur at $\Gamma$, as is the case when doping away from charge neutrality.
We now make the assumption that the quasiparticle effective mass is considerably lighter than the Goldstone mass. Then a good basis of states, for small $\bf{q}$, should be the quasiparticle at $\Gamma$ and the Goldstone at $\bf{q}$. Hence, we perform perturbation theory in the states $\ket{\Phi^{(e_Y',e_Y)}_{\bs{p};\bs{0},\alpha}}$, which are eigenstates at $\Gamma$ but not away from it.
Furthermore, we approximate the Goldstone modes entering Eq.~\eqref{eqmain: varbasis} by their form at $\Gamma$, that is, $G^{(e_Y)}_{\bs{p}-\bs{q},\bs{k}} \approx 1/\sqrt{N_{\mathcal{M}}}$, to obtain the approximation
\begin{equation} \label{eqmain: perturbedH}
\begin{aligned}
&\tilde{\mathcal{H}}^{(\bs{p},e_Y',e_Y)}_{\pm} = \frac{\mathcal{A}^{(\bs{p},e_Y',e_Y)}_{\bs{0},\bs{0}} \pm \delta_{e_Y,e_Y'} \mathcal{B}^{(\bs{p},e_Y)}_{\bs{0},\bs{0}}}{1 \pm \delta_{e_Y,e_Y'}/N_{\mathcal{M}}}, \\
&\mathcal{A}_{\bs{0},\bs{0}}^{(\bs{p},e_Y',e_Y)} = R(\bs{0}) + \epsilon(\bs{p}), \\
&\mathcal{B}_{\bs{0},\bs{0}}^{(\bs{p},e_Y)} = \frac{1}{N_{\mathcal{M}}} \bigg\{2R(\bs{0}) + R(-\bs{p}) \\
&+2 \sum_{\bs{k}} \bigg[S^{(e_Y,e_Y)*}_{\bs{k};\bs{0}} (\bs{0}) - S^{(e_Y,e_Y)}_{\bs{k}-\bs{p};-\bs{p}} (\bs{p})
- S^{(e_Y,e_Y)}_{-\bs{p};\bs{k}-\bs{p}} (\bs{p}) 
\bigg]
\bigg\}.
\end{aligned}
\end{equation}
Here, we have also used that the Hamiltonian in Eq.~\eqref{eqmain: chiral_limit_trions} becomes block-diagonal in the basis $(\ket{\Phi^{(e_Y',e_Y)}_{\bs{p};\bs{q},1}} \pm \ket{\Phi^{(e_Y',e_Y)}_{\bs{p};\bs{q},2}})/\sqrt{2}$, corresponding to the two blocks $\tilde{\mathcal{H}}^{(\bs{p},e_Y',e_Y)}_{\pm}$ in Eq.~\eqref{eqmain: perturbedH}.
Importantly, the norm of the matrix $N_{\mathcal{M}} \mathcal{B}^{(\bs{p},e_Y)}$ does not grow with $N_{\mathcal{M}}$. Hence, in the thermodynamic limit we obtain
$\tilde{\mathcal{H}}^{(\bs{p},e_Y',e_Y)}_{\pm} \stackrel{N_{\mathcal{M}} \rightarrow \infty}{=} R(\bs{0}) + \epsilon(\bs{p})$,
which is part of the particle-Goldstone continuum and so there is no bound state at $\Gamma$. For the special case $\nu=-1, Q=1$ (Fig.~\ref{figmain: projectedspectra}\textbf{a}), this argument fails because the $Q=1$ dispersion is anomalously flat~\cite{TBG5, VafekKangPRB2021, song2021matbg} (observation explained by the heavy fermion model of TBG in \cite{song2021matbg}) and so the basis of states is not justified. 

\begin{figure}[t]
\includegraphics[width=0.48\textwidth]{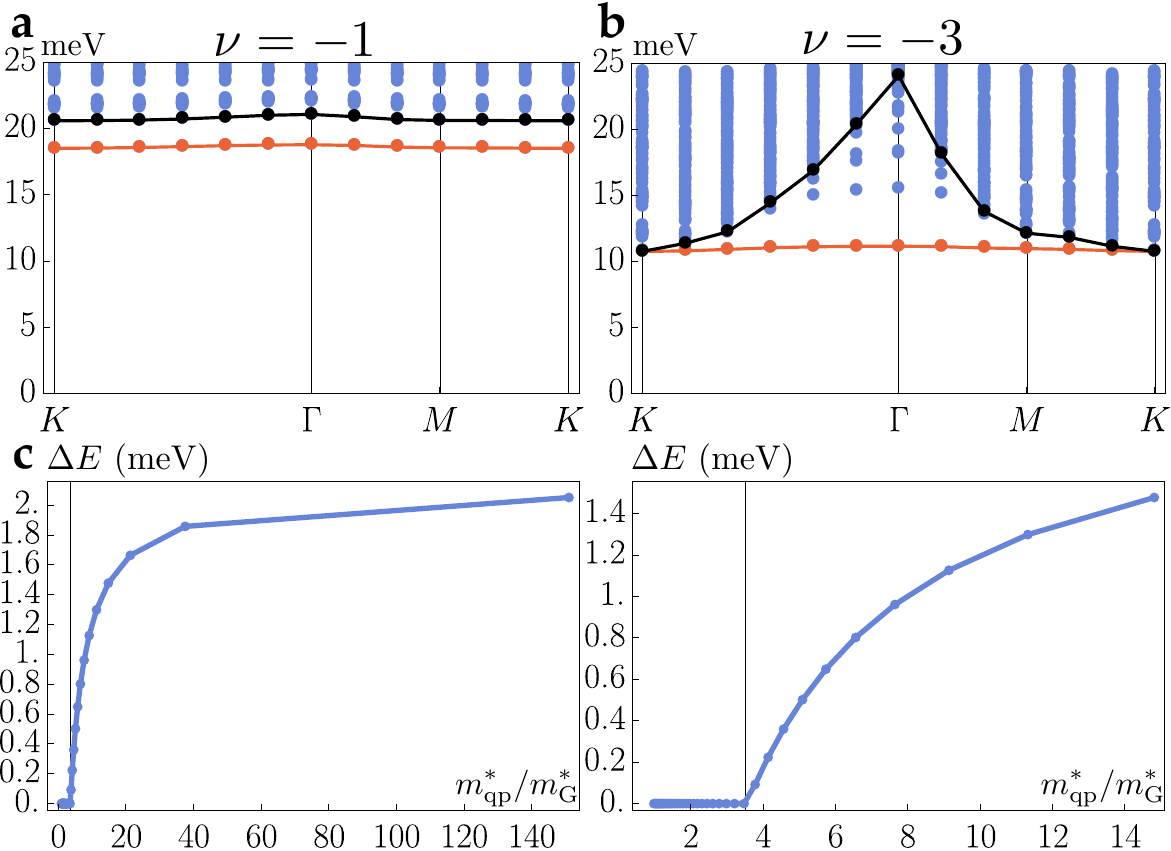}
\caption{Exact $N=3, Q=1$ spectra with minimum of quasiparticle-Goldstone continuum highlighted (black) for fillings $\nu=-1,-3$ in the chiral limit. See Fig.~\ref{figmain: exactspectra} for model parameters. \textbf{a}~For $\nu=-1, Q=1$, the trion bound state persists at all momenta (for the chiral limit). \textbf{b}~For $\nu=-3, Q=1$, a bound state is only present away from the $K$ point. \textbf{c}~Dependence of the binding energy $\Delta E$ on the ratio of quasiparticle and Goldstone masses (see also Tab.~\ref{tabmain: effectivemasses}). For this plot, we have artificially tuned the $\nu=-1$, $N=Q=1$ quasiparticle dispersion away from its flatband limit, thereby gradually reducing $m^*_{\mathrm{qp}}$. The transition mass ratio above which a bound state exists is estimated as $m^*_{\mathrm{qp}}/m^*_{\mathrm{G}} \approx 3.5$. We have restricted our binding energy analysis to $\nu=-1$ because no other filling exhibits a well-separated bound state at all momenta (Fig.~\ref{figmain: exactspectra}).}
\label{figmain: exactchiralspectra}
\end{figure}

Fig.~\ref{figmain: exactchiralspectra}\textbf{c} shows that the trion binding energy increases monotonically with the ratio $m^*_{\mathrm{qp}}/m^*_{\mathrm{G}}$ of the quasiparticle effective mass and the Goldstone mass $m^*_{\mathrm{G}}$. Hence, the trion binding energy follows directly from this ratio, which is tabulated in Tab.~\ref{tabmain: effectivemasses} for all choices of $\nu$ and $Q$. Only for $m^*_{\mathrm{qp}}/m^*_{\mathrm{G}} > 3.5$ is there a trion bound state across the entire moir\'e BZ, which is satisfied for $(\nu,Q) = (-1,1)$ and $(\nu,Q) = (-2,1)$. Moreover, since the mass ratio $m^*_{\mathrm{qp}}/m^*_{\mathrm{G}} = 4.28$ for $(\nu,Q) = (-2,1)$ is rather close to the critical value, the corresponding bound state has a very small binding energy ($\sim 0.5 \mathrm{meV}$) and we expect it to become unbound away from the chiral limit. In fact, and in contrast to the bound state at $(\nu,Q) = (-1,1)$, this bound state is absent for screening length $\xi=10$nm, consistent with a $(\nu,Q) = (-2,1)$ mass ratio $m^*_{\mathrm{qp}}/m^*_{\mathrm{G}} = 3.3$ at this screening length.

\section{Presence of bound state away from the particle-hole minimum and variational wavefunction}
The presence of a trion bound state is intimately tied to the lowest band of the inverse overlap matrix $[\mathcal{O}^{(\bs{p},e_Y',e_Y)}]^{-1}$ in Eq.~\eqref{eqmain: chiral_limit_trions}. This band is approximately flat, for $\nu=0$ it appears at eigenvalue $\sim 0.77$ and yields a $\mathcal{H}^{(\bs{p},e_Y',e_Y)}$ expectation value $\sim 35.8 \mathrm{meV}$, explaining the presence of a flat trion bound state at energy $\sim 0.77 \times 35.8 \mathrm{meV} = 27.6 \mathrm{meV}$ in the physical spectrum of $\tilde{\mathcal{H}}^{(\bs{p},e_Y',e_Y)}$ (Fig.~\ref{figmain: goldstoneprojection}\textbf{d}) (the quasiparticle-Goldstone continuum begins at $\sim 30 \mathrm{meV}$). In fact, the actual bound state energy is slightly lower ($26.1 \mathrm{meV}$), because the low $[\mathcal{O}^{(\bs{p},e_Y',e_Y)}]^{-1}$ band is not an exact eigenstate of $\tilde{\mathcal{H}}^{(\bs{p},e_Y',e_Y)}$. The resulting flat bound state merges with the continuum when the quasiparticle energy dips below $26.1 \mathrm{meV}$ close to $\Gamma$. 
Conversely, the bound state at $\nu=-1$, $Q=1$ is already present at low energies in $\mathcal{H}^{(\bs{p},e_Y',e_Y)}$, and has $>99\%$ overlap with the low $[\mathcal{O}^{(\bs{p},e_Y',e_Y)}]^{-1}$ flatband, which only reduces its binding energy in the physical spectrum of $\tilde{\mathcal{H}}^{(\bs{p},e_Y',e_Y)}$.
We find that for all choices of $\nu$, $Q$, the lowest trion bound state always has very high overlap with a low-eigenvalue flat band of $[\mathcal{O}^{(\bs{p},e_Y',e_Y)}]^{-1}$, and so capitalizes on the non-orthogonality between the variational states of Eq.~\eqref{eqmain: varbasis}.

For the $\nu=0$, $Q=1$ trion bound state that develops away from $\Gamma$ in Fig.~\ref{figmain: goldstoneprojection}\textbf{d}, we find good ($\geq90\%$) overlap with the variational state
\begin{equation}
    |\mathrm{TBS}_{\bs{p}}\rangle = \frac{1}{\sqrt{2}}\sum_{\bs{q}} \phi^{\bs{p}}_{\bs{q}} \left(\ket{\Phi^{(e_Y,e_Y)}_{\bs{p};\bs{q},1}} - \ket{\Phi^{(e_Y,e_Y)}_{\bs{p};\bs{q},2}}\right),
\end{equation}
\begin{equation}
\begin{aligned}
\Phi^{\bs{p}}_{\bs{q}} = \frac{e^{-\mathrm{i}\arg G_{\bs{p}-\bs{q},\bs{p}}}}{Z_{\bs{p}}} \sum_{\bs{G} \in Q_0}& \bigg[\exp\left(-2\frac{|\bs{q}|^2 +|\bs{p}-\bs{q}+\bs{G}|^2}{|\bs{p}|^2}\right)\\
&\times \exp\left({\mathrm{i}\angle[\bs{p}-\bs{q}+\bs{G}]}\right)\bigg],
\end{aligned}
\end{equation}
where $\angle[\bs{p}-\bs{q}+\bs{G}]$ denotes the angle of the vector $\bs{p}-\bs{q}+\bs{G}$ to the horizontal, $Z_{\bs{p}}$ is a normalization factor, and $\arg G_{\bs{p}-\bs{q},\bs{p}}$ is the phase of the Goldstone wavefunction $G_{\bs{p}-\bs{q},\bs{p}}$. Note that a phase factor of the form $e^{-\mathrm{i}\arg G_{\bs{p}-\bs{q},\bs{q}'}}$ must enter in $\Phi^{\bs{p}}_{\bs{q}}$ to account for gauge invariance under multiplying the individual Goldstone eigenstates $G^{(e_Y)}_{\bs{p}-\bs{q},\bs{k}}$ by a phase in Eq.~\eqref{eqmain: varbasis}. However, the choice $\bs{q}'=\bs{p}$ is particular to the trion bound state.

\section{Conclusions}
The present analysis allows us to study the relative stability of the single particle excitations found in~\cite{TBG5,VafekKangPRL2020,KangBernevigVafek2021,VafekKangPRB2021} and the bound trion excitations whose first example was shown to exist in TBG in \cite{TBG6}. On the light mass side bound trions do not compete with the lowest energy single particle excitations (they only become competitive at large momenta away from the band minimum), reinforcing the notion of the light Fermi liquid advanced in Ref.~\onlinecite{KangBernevigVafek2021}. On the heavy mass side the bound trion states can be energetically competitive, and even if they are not the absolute lowest energy excitations (except on the heavy mass side of the idealized chiral limit at $|\nu|=1$), for a range of their total momentum, they can form below the quasiparticle-Goldstone continuum, further reinforcing the heavy-light dichotomy~\cite{KangBernevigVafek2021}.

Despite not being the lowest energy charged excitations, even on the light mass side the bound trion states can appear as higher energy isolated states below the quasiparticle-Goldstone continuum. We expect this to have an imprint in the STM spectroscopy at a fixed filling, which we plan to elucidate in a future publication. 

\begin{acknowledgments}
We thank F. Xie, S. Parameswaran and Y. Kwan for discussions. F.S. was supported by a fellowship at the Princeton Center for Theoretical Science.  O.V. was supported by the NSF Grant No. DMR-1916958 and, in part, by the NSF Grant No. DMR-1644779 and the state of Florida MagLab core grant. This work also received support from the DOE Grant No. DE-SC0016239, the Schmidt Fund for Innovative Research, Simons Investigator Grant No. 404513, the Packard Foundation, the Gordon and Betty Moore Foundation through Grant No. GBMF8685 towards the Princeton theory program, and a Guggenheim Fellowship from the John Simon Guggenheim Memorial Foundation. Further support was provided by the  NSF-MRSEC Grant No. DMR-2011750, ONR Grant No. N00014-20-1-2303, BSF Israel US foundation Grant No. 2018226, and the Princeton Global Network Funds.

\emph{Note added: } During the final stages of preparation of this manuscript, trion bound states were predicted~\cite{khalaf2021electrons} for all fillings $\nu\leq -1$ at screening length $\xi=100$nm~\cite{EslamEmail}, Fig. 3\textbf{a} of Ref.~\onlinecite{khalaf2021electrons}, using an approximate model of TBG, which neglects the mixing between different Chern number sectors away from the chiral limit. Instead, our model, which takes the mixing of Chern number sectors into account, and for which we use realistic screening lengths of $\xi\leq20$nm~\cite{cao_correlated_2018,lu2020fingerprints,stepanov_interplay_2020}, predicts a gapped trion bound state only near the chiral limit and at $\nu=-1$.
\end{acknowledgments}

\widetext
\appendix
\section{Review of notation}
We begin the appendices by reviewing our notation, which follows that of Ref.~\onlinecite{TBG5}.

\subsection{Units and conventions} \label{sec: units_and_values}
Using the graphene Fermi velocity $v_{\mathrm{F}} = 5.944 \mathrm{eV}$\AA\ and Bistritzer-MacDonald (BM)~\cite{Bistritzer11} continuum model hopping strength $w_1 = 110 \mathrm{meV}$~\cite{TBG1}, we find $w_1/v_{\mathrm{F}} k_\theta = 0.593$, where $k_\theta = 2 |\bs{K}| \sin{\frac{\theta}{2}}$ is the separation of the Dirac cones of the two twisted bilayer graphene (TBG) layers. Here, $\bs{K}$ is a high-symmetry momentum (the $K$ point) in the Brillouin zone of the graphene monolayer, and $\theta$ is the twist angle of TBG.
The unit cell area $A_{\mathrm{UC}}$ and the Brillouin zone area $A_{\mathrm{BZ}} = k_\theta^2 3\sqrt{3}/2$ of the TBG Moiré lattice are related by $A_{\mathrm{UC}} = 4 \pi^2/A_{\mathrm{BZ}}$. Hence, the total sample area $\Omega$ of TBG is related to the number of momenta $N_{\mathcal{M}}$ in the Moiré Brillouin zone by
\begin{equation} \label{eq: moirearea}
\Omega = \frac{8 \pi^2 N_{\mathcal{M}}}{3 \sqrt{3} k_\theta^2}.
\end{equation}
In practice, we use $|\bs{K}| = 1.703$\AA$^{-1}$\ and $\theta = 1.05^\circ$.
In all plots, we use the double gate screened Coulomb potential unless otherwise stated~\cite{TBG3}:
\begin{equation} \label{eq: coulombdoublegate}
V(\bs{q}) = \pi \xi^2 U_\xi \frac{\tanh (\xi |\bs{q}|/2)}{\xi |\bs{q}|/2},
\end{equation}
where the screening length $\xi$ and the associated energy scale $U_\xi$ are defined in Ref.~\onlinecite{TBG3}. In practice, we consider two regimes: (1) $\xi = 10 \mathrm{nm}$, $U_\xi = 26 \mathrm{meV}$; and (2) $\xi = 20 \mathrm{nm}$, $U_\xi = 13 \mathrm{meV}$.

\subsection{Flat-band projected Hamiltonian}
We assume a perfectly flat dispersion (nonchiral-flat limit), so that the full Hamiltonian of twisted bilayer graphene (TBG) is a flatband-projected positive-semidefinite Hamiltonian (PSDH)~\cite{KangVafek19, Bultinck2020PRX, TBG3}:
\begin{equation} \label{eq: nonchiral-flat Hamiltonian}
H = \frac{1}{2\Omega} \sum_{\bs{q},\bs{G}} O^\dagger_{\bs{q}, \bs{G}} O_{\bs{q},\bs{G}},
\quad O_{\bs{q}, \bs{G}} = \sqrt{V(\bs{q}+\bs{G})} \sum_{\bs{k},m,n,\eta,s} M^{(\eta)}_{m,n} (\bs{k},\bs{q}+\bs{G}) \left(c^\dagger_{\bs{k} + \bs{q},m,\eta,s} c_{\bs{k},n,\eta,s} - \frac{1}{2} \delta_{\bs{q},\bs{0}} \delta_{m,n} \right).
\end{equation}
Here, $\bs{q}$ ranges over the Moiré Brillouin zone (mBZ), whose number of discrete elements is fixed by the total area $\Omega$ of the TBG sample via Eq.~\eqref{eq: moirearea}, and $\bs{G}$ ranges over the Moiré reciprocal lattice $\mathcal{Q}_0$ defined in Ref.~\onlinecite{TBG1}. Furthermore, $c^\dagger_{\bs{k},n,\eta,s}$ creates an electron at Moiré momentum $\bs{k}$ in the eigenstate $n=\pm$ of the BM continuum Hamiltonian for graphene valley $\eta = \pm$ and spin $s = \pm$. We thereby restrict to the 8 BM bands closest to the Fermi level, these are anomalously flat at the first magic angle, justifying the absence of a kinetic term in $H$. $V(\bs{q})$ is the Fourier-transformed Coulomb interaction, given in Eq.~\eqref{eq: coulombdoublegate}. Moreover, the form factors are defined as
\begin{equation} \label{eq: formfactors}
M^{(\eta)}_{m,n} (\bs{k},\bs{q}+\bs{G}) = \sum_{\alpha, \bs{Q}} u^*_{\bs{Q}-\bs{G},\alpha;m,\eta}(\bs{k} + \bs{q}) u_{\bs{Q},\alpha;n,\eta}(\bs{k}),
\end{equation}
where $\alpha$ ranges over the two graphene sublattices $A$ and $B$, $\bs{Q}$ ranges over the hexagonal lattice $\mathcal{Q}_\pm$ defined in Ref.~\onlinecite{TBG1}, and $u_{\bs{Q},\alpha;n,\eta}(\bs{k})$ are the BM Bloch states. Due to Moiré periodicity, these satisfy 
\begin{equation}
u_{\bs{Q},\alpha;n,\eta} (\bs{k} + \bs{G}) = u_{\bs{Q}-\bs{G},\alpha;n,\eta} (\bs{k}).
\end{equation}
Moreover, the form factors satisfy
\begin{equation}
M^{(\eta)*}_{m,n}(\bs{k},\bs{q}+\bs{G}) = M^{(\eta)}_{n,m}(\bs{k}+\bs{q},-\bs{q}-\bs{G}).
\end{equation}

\subsection{Ground states} \label{sec: groundstates}
We next derive the ground state (candidates) of TBG at even integer filling.
The Hamiltonian $H$ in Eq.~\eqref{eq: nonchiral-flat Hamiltonian} is manifestly positive-semidefinite. At half filling (charge neutrality, $\nu = 0$), one possible ground state is the intervalley-coherent Slater determinant state~\cite{KangVafek19,Bultinck2020PRX, TBG4}
\begin{equation} \label{eq: defgroundstate}
\ket{\Psi_0} = \prod_{\bs{k}} \left(c^\dagger_{\bs{k},+,+,+} c^\dagger_{\bs{k},-,+,+} c^\dagger_{\bs{k},+,-,-} c^\dagger_{\bs{k},-,-,-}\right) \ket{0}, \quad O_{\bs{q}, \bs{G}} \ket{\Psi_0} = 0,
\end{equation}
where we have used that the form factors $M^{(\eta)}_{m,n} (\bs{k},\bs{q}+\bs{G})$ do not depend on spin in order to derive the second equality. This state is a $U(4)$ ferromagnet~\cite{KangVafek19, Bultinck2020PRX, TBG4} and therefore supports gapless Goldstone modes. Further degenerate ground states can be obtained by changing the valley-spin occupation in Eq.~\eqref{eq: defgroundstate} to other configurations involving opposite valley indices. Without loss of generality, we will fix $\ket{\Psi_0}$ as ground state at charge neutrality in the following.
At filling $\nu = -2$, one ground state candidate is the intervalley-coherent Slater determinant state~\cite{KangVafek19, Bultinck2020PRX, TBG4}
\begin{equation} \label{eq: defgroundstatenm2}
\ket{\Psi_{-2}} = \prod_{\bs{k}} \left(c^\dagger_{\bs{k},+,+,+} c^\dagger_{\bs{k},-,+,+}\right) \ket{0}, \quad O_{\bs{q}, \bs{G}} \ket{\Psi_{-2}} = \delta_{\bs{q},\bs{0}} E_{\bs{G}} \ket{\Psi_{-2}}, \quad E_{\bs{G}} = - \sqrt{V(\bs{G})} \sum_{\bs{k},m} M^{(-)}_{m,m} (\bs{k},\bs{G}).
\end{equation}
Similar to $\ket{\Psi_0}$, the state $\ket{\Psi_-2}$ is a $U(4)$-ferromagnetic state and supports Goldstone mode excitations. Refs ~\onlinecite{KangVafek19,TBG4} showed that $\ket{\Psi_-2}$ is an exact ground state of $H-\mu N$, with suitably chosen chemical potential $\mu$, when the \emph{flat metric condition} is fulfilled, which posits that the form factors [Eq.~\eqref{eq: formfactors}] are independent of $\bs{k}$ at $\bs{q}=\bs{0}$. Here, the operator $N$ measures the particle number from the point of charge neutrality. Without the flat metric condition -- which does not hold exactly for TBG -- there is no guarantee that $\ket{\Psi_{-2}}$ is a ground state, however, it is always an exact eigenstate \cite{KangVafek19, Bultinck2020PRX, TBG4}. A necessary condition for $\ket{\Psi_{-2}}$ to be a ground state is that we can find a chemical potential $\mu$ so that both the charge $+1$ and $-1$ gaps are finite. In the chiral limit only, we can furthermore derive ground states for fillings $\nu=-1$ and $\nu=-3$~\cite{TBG3, VafekKangPRB2021,kwan2021kekule, Bultinck2020PRX}.

\subsection{General excitation spectra}
We next describe the general method \cite{KangVafek19, TBG5, VafekKangPRB2021} to find the spectra of excitations above the ground state (candidates) of the TBG PSDH that were derived in Sec.~\ref{sec: groundstates}.
The ground state $\ket{\Psi_0}$ at half filling, Eq.~\eqref{eq: defgroundstate}, satisfies 
\begin{equation} 
O_{\bs{q}, \bs{G}} \ket{\Psi_0} = 0.
\end{equation}
For an excitation operator $\mathcal{E}$, we then find that
\begin{equation} \label{eq: generalnu0scatteringeq}
H \mathcal{E} \ket{\Psi} = \frac{1}{2\Omega} \sum_{\bs{q},\bs{G}} \left[O_{-\bs{q},-\bs{G}}, \left[O_{\bs{q},\bs{G}},\mathcal{E} \right] \right]\ket{\Psi},
\end{equation}
so that we only need to evaluate the (double) commutator $\left[O_{-\bs{q},-\bs{G}}, \left[O_{\bs{q},\bs{G}},\mathcal{E} \right] \right]$ in order to find the scattering matrix for all excitations mixing with $\mathcal{E}\ket{\Psi_0}$. In general, these will be all states $\mathcal{E}'\ket{\Psi_0}$ where $\mathcal{E}'$ has the same charge and total momentum quantum numbers as $\mathcal{E}$.
Ground state (candidates) $\ket{\Psi}$ at finite filling satisfy 
\begin{equation} \label{eq: genericfinitefillingOeigstate}
O_{\bs{q}, \bs{G}} \ket{\Psi} = \delta_{\bs{q},\bs{0}} E_{\bs{G}} \ket{\Psi},
\end{equation}
see for example Eq.~\eqref{eq: defgroundstatenm2} for filling $\nu = -2$. For an excitation operator $\mathcal{E}$, we then find that
\begin{equation} \label{eq: genericscattmat}
(H-\mu N) \mathcal{E} \ket{\Psi} = \left\{ \frac{1}{2\Omega} \sum_{\bs{q},\bs{G}} \left[O_{-\bs{q},-\bs{G}}, \left[O_{\bs{q},\bs{G}},\mathcal{E} \right] \right] + \frac{1}{\Omega} \sum_{\bs{G}} E_{-\bs{G}} \left[O_{\bs{0},\bs{G}},\mathcal{E}\right] + \left[\frac{1}{2\Omega} \left(\sum_{\bs{G}} E_{-\bs{G}} E_{\bs{G}}\right)-\mu N \right] \mathcal{E}\right\}\ket{\Psi}.
\end{equation}
Here, the first term is the same as in Eq.~\eqref{eq: generalnu0scatteringeq}. The second ``Hartree" term modifies the single-particle dispersion away from charge neutrality. In particular, and unlike for $\nu=0$, it yields different single-particle and single-hole spectra. The third term represents a particle number-dependent energy shift, where $\mu$ must be tuned to stabilize $\ket{\Psi}$ as a ground state, so that both the charge $+1$ and $-1$ gaps are finite.

\section{Scattering matrices}
We next derive the excitation scattering matrix in different quantum number sectors, where we denote the total number of particles by $N$ and the total charge by $Q$. We define $N$ so that it counts the number of creation \emph{and} annihilation operators of a given excitation. This operator is \emph{not} associated with a symmetry of the PSDH in Eq.~\eqref{eq: nonchiral-flat Hamiltonian}. Instead, $N$ emerges as a conserved quantity of excited state scattering matrices because the TBG ground states are U(4) ferromagnets: these states have a given set of valley-spin flavors fully occupied, and so do not support particle-hole excitations that would change $N$ while preserving $Q$ as well as valley and spin quantum numbers. On the other hand, the conservation of $Q$ follows from the global U(1) gauge symmetry of the PSDH in Eq.~\eqref{eq: nonchiral-flat Hamiltonian}. As a result of $N$-conservation in this Krylov subspace, the scattering equation maintains the particle number subspace. 

\subsection{$N=1$, $Q=1$}
We begin by deriving the single-particle excitations above the ground states of Sec.~\ref{sec: groundstates}. At each total momentum $\bs{k}$, these can be obtained by diagonalizing a $2 \times 2$ matrix \cite{VafekKangPRB2021,TBG5}.
For the charge $+1$ operator $\mathcal{E} = c^\dagger_{\bs{k},m,\eta,s}$, we obtain
\begin{equation}
H c^\dagger_{\bs{k},m,\eta,s} \ket{\Psi_0} = \sum_{n} R^{(\eta)}_{n,m} (\bs{k}) c^\dagger_{\bs{k},n,\eta,s} \ket{\Psi_0}.
\end{equation}
Here, we have defined the matrices
\begin{equation} \label{eq: rmatdef}
R^{(\eta)}_{n,m} (\bs{k}) = \frac{1}{2\Omega} \sum_{\bs{q},\bs{G}} V(\bs{q}+\bs{G}) \sum_{l} M^{(\eta)*}_{l,n}(\bs{k},\bs{q}+\bs{G}) M^{(\eta)}_{l,m} (\bs{k},\bs{q}+\bs{G}),
\end{equation}
whose spectrum yields the single-particle dispersion.
For a ground state $\ket{\Psi}$ at finite filling, we find by using Eq.~\eqref{eq: genericscattmat} that 
\begin{equation}
\left[H, c^\dagger_{\bs{k},m,\eta,s}\right] \ket{\Psi} = \sum_{n} \left[R^{(\eta)}_{n,m} (\bs{k}) + \frac{1}{\Omega} \sum_{\bs{G}} E_{-\bs{G}} 
\sqrt{V(\bs{G})} M^{(\eta)}_{n,m}(\bs{k},\bs{G})
\right] c^\dagger_{\bs{k},n,\eta,s} \ket{\Psi}.
\end{equation}
For simplicity, here we only give the commutator. The full expression follows from adding the third term of Eq.~\eqref{eq: genericscattmat}. For future reference, we define \begin{equation} \label{eq: finitefillingRmatplus}
\tilde{R}^{(\eta,+)}_{n,m} (\bs{k},E_{\bs{G}} ) = R^{(\eta)}_{n,m} (\bs{k}) + \frac{1}{\Omega} \sum_{\bs{G}} E_{-\bs{G}} 
\sqrt{V(\bs{G})} M^{(\eta)}_{n,m}(\bs{k},\bs{G}),
\end{equation}
which is a functional of $E_{\bs{G}}$ as defined by Eq.~\eqref{eq: genericfinitefillingOeigstate}.

\subsection{$N=1$, $Q=-1$}
Correspondingly, for the charge $-1$ operator $\mathcal{E} = c_{\bs{k},m,\eta,s}$, we obtain \cite{VafekKangPRB2021,TBG5}
\begin{equation}
H c_{\bs{k},m,\eta,s} \ket{\Psi_0} = \sum_{n} R^{(\eta)}_{m,n} (\bs{k}) c_{\bs{k},n,\eta,s} \ket{\Psi_0}.
\end{equation}
We hence find that the single-particle charge $+1$ and charge $-1$ excitation spectra are identical at charge neutrality.
For a ground state candidate $\ket{\Psi}$ at finite filling, we find by using Eq.~\eqref{eq: genericscattmat} that 
\begin{equation}
\left[H, c_{\bs{k},m,\eta,s}\right] \ket{\Psi} = \sum_{n} \left[R^{(\eta)}_{m,n} (\bs{k}) - \frac{1}{\Omega} \sum_{\bs{G}} E_{-\bs{G}} 
\sqrt{V(\bs{G})} M^{(\eta)}_{m,n}(\bs{k},\bs{G})
\right] c_{\bs{k},n,\eta,s} \ket{\Psi},
\end{equation}
again up to a total energy shift [third term of Eq.~\eqref{eq: genericscattmat}] that can be offset by a suitable choice of chemical potential $\mu$.
For future reference, we define 
\begin{equation} \label{eq: finitefillingRmatminus}
\tilde{R}^{(\eta,-)}_{m,n} (\bs{k},E_{\bs{G}} ) = R^{(\eta)}_{m,n} (\bs{k}) - \frac{1}{\Omega} \sum_{\bs{G}} E_{-\bs{G}} 
\sqrt{V(\bs{G})} M^{(\eta)}_{m,n}(\bs{k},\bs{G}).
\end{equation}

\subsection{$N=2$, $Q=2$} \label{sec: n2q2scattmat}
We next derive the two-particle excitations above $\ket{\Psi_0}$, first presented in \cite{TBG5}. At each total momentum $\bs{p}$, these can be obtained by diagonalizing a $4N_{\mathcal{M}} \times 4N_{\mathcal{M}}$ matrix, where $N_{\mathcal{M}}$ is the number of momenta in the Moiré Brillouin zone as set by the total sample area $\Omega$ via Eq.~\eqref{eq: moirearea}.
For the charge $+2$ operator $\mathcal{E} = c^\dagger_{\bs{k}+\bs{p},m_1,\eta_1,s_1} c^\dagger_{-\bs{k},m_2,\eta_2,s_2}$, we obtain
\begin{equation} \label{eq: duonscattmat}
\begin{aligned}
H c^\dagger_{\bs{k}+\bs{p},m_1,\eta_1,s_1} c^\dagger_{-\bs{k},m_2,\eta_2,s_2} \ket{\Psi_0} = \sum_{\tilde{\bs{k}},\tilde{m}_1,\tilde{m}_2} \bigg\{&\delta_{\bs{k},\tilde{\bs{k}}} \left[R^{(\eta_1)}_{\tilde{m}_1,m_1} (\bs{k}+\bs{p})\delta_{\tilde{m}_2,m_2} + R^{(\eta_2)}_{\tilde{m}_2,m_2} (-\bs{k})\delta_{\tilde{m}_1,m_1} \right] \\&+2T^{(\eta_1,\eta_2)}_{\tilde{\bs{k}},\tilde{m}_1,\tilde{m}_2;\bs{k},m_1,m_2} (\bs{p}) \bigg\} c^\dagger_{\tilde{\bs{k}}+\bs{p},\tilde{m}_1,\eta_1,s_1} c^\dagger_{-\tilde{\bs{k}},\tilde{m}_2,\eta_2,s_2} \ket{\Psi_0},
\end{aligned}
\end{equation}
where we have defined the electron-electron scattering matrix
\begin{equation} \label{eq: charge2TMAT}
T^{(\eta_1,\eta_2)}_{\tilde{\bs{k}},\tilde{m}_1,\tilde{m}_2;\bs{k},m_1,m_2} (\bs{p}) = \frac{1}{2\Omega} \sum_{\bs{G}} V(\bs{k}-\tilde{\bs{k}}+\bs{G}) M^{(\eta_1)*}_{m_1,\tilde{m}_1} (\tilde{\bs{k}}+\bs{p},\bs{k}-\tilde{\bs{k}}+\bs{G}) M^{(\eta_2)}_{\tilde{m}_2,m_2} (-\bs{k},\bs{k}-\tilde{\bs{k}}+\bs{G}).
\end{equation}
In this expression, the momentum $(\bs{k}-\tilde{\bs{k}})$ must be treated as an element of the first Moiré Brillouin zone, dropping any offsets by reciprocal lattice vectors. For an infinite $\mathcal{Q}_0$ lattice, this convention has no effect, however, it is required for numerical consistency when doing simulations on a finite $\mathcal{Q}_0$ lattice.
The corresponding scattering matrix for finite filling -- modulo the third, diagonal term in Eq.~\eqref{eq: genericscattmat} -- is obtained from Eq.~\eqref{eq: duonscattmat} via the replacement
\begin{equation}
R^{(\eta)}_{n,m} (\bs{k}) \rightarrow \tilde{R}^{(\eta,+)}_{n,m} (\bs{k},E_{\bs{G}}),
\end{equation}
where $\tilde{R}^{(\eta,+)}_{n,m} (\bs{k},E_{\bs{G}})$ is defined in Eq.~\eqref{eq: finitefillingRmatplus}.

\subsection{$N=2$, $Q=0$} \label{sec: excitons}
For the charge $0$ operator $\mathcal{E} = c^\dagger_{\bs{k}+\bs{p},m_1,\eta_1,s_1} c_{\bs{k},m_2,\eta_2,s_2}$, refs \cite{TBG5,VafekKangPRB2021} obtained:
\begin{equation} \label{eq: excitonscattmat}
\begin{aligned}
H c^\dagger_{\bs{k}+\bs{p},m_1,\eta_1,s_1} c_{\bs{k},m_2,\eta_2,s_2} \ket{\Psi_0} = \sum_{\tilde{\bs{k}},\tilde{m}_1,\tilde{m}_2} \bigg\{&\delta_{\bs{k},\tilde{\bs{k}}} \left[R^{(\eta_1)}_{\tilde{m}_1,m_1} (\bs{k}+\bs{p})\delta_{\tilde{m}_2,m_2} + R^{(\eta_2)}_{m_2,\tilde{m}_2} (\bs{k})\delta_{\tilde{m}_1,m_1} \right] \\&-2S^{(\eta_1,\eta_2)}_{\tilde{\bs{k}},\tilde{m}_1,\tilde{m}_2;\bs{k},m_1,m_2} (\bs{p}) \bigg\} c^\dagger_{\tilde{\bs{k}}+\bs{p},\tilde{m}_1,\eta_1,s_1} c_{\tilde{\bs{k}},\tilde{m}_2,\eta_2,s_2} \ket{\Psi_0},
\end{aligned}
\end{equation}
where we have defined the electron-hole scattering matrix
\begin{equation} \label{eq: neutralscattmat}
S^{(\eta_1,\eta_2)}_{\tilde{\bs{k}},\tilde{m}_1,\tilde{m}_2;\bs{k},m_1,m_2} (\bs{p}) = \frac{1}{2\Omega} \sum_{\bs{G}} V(\bs{k}-\tilde{\bs{k}}+\bs{G}) M^{(\eta_1)*}_{m_1,\tilde{m}_1} (\tilde{\bs{k}}+\bs{p},\bs{k}-\tilde{\bs{k}}+\bs{G}) M^{(\eta_2)}_{m_2,\tilde{m}_2} (\tilde{\bs{k}},\bs{k}-\tilde{\bs{k}}+\bs{G}).
\end{equation}
In this expression, the momentum $(\bs{k}-\tilde{\bs{k}})$ must again be treated as an element of the first Moiré Brillouin zone.
The corresponding scattering matrix for finite filling -- except for the third, diagonal term in Eq.~\eqref{eq: genericscattmat} -- is obtained from Eq.~\eqref{eq: excitonscattmat} via the replacements
\begin{equation}
R^{(\eta_1)}_{\tilde{m}_1,m_1} (\bs{k}+\bs{p}) \rightarrow \tilde{R}^{(\eta_1,+)}_{\tilde{m}_1,m_1} (\bs{k}+\bs{p},E_{\bs{G}}), \quad 
R^{(\eta_2)}_{m_2,\tilde{m}_2} (\bs{k}) \rightarrow R^{(\eta_2,-)}_{m_2,\tilde{m}_2} (\bs{k},E_{\bs{G}}),
\end{equation}
where $\tilde{R}^{(\eta,+)}_{n,m} (\bs{k},E_{\bs{G}})$ and $\tilde{R}^{(\eta,-)}_{n,m} (\bs{k},E_{\bs{G}})$ are defined in Eqs.~\eqref{eq: finitefillingRmatplus} and~\eqref{eq: finitefillingRmatminus}, respectively.
Finally, we note that the charge $0$ scattering matrix $S^{(\eta_1,\eta_2)}_{\tilde{\bs{k}},\tilde{m}_1,\tilde{m}_2;\bs{k},m_1,m_2} (\bs{p})$ of Eq.~\eqref{eq: neutralscattmat} and the charge $+2$ scattering matrix $T^{(\eta_1,\eta_2)}_{\tilde{\bs{k}},\tilde{m}_1,\tilde{m}_2;\bs{k},m_1,m_2} (\bs{p})$ of Eq.~\eqref{eq: charge2TMAT} are related by
\begin{equation}
T^{(\eta_1,\eta_2)}_{\tilde{\bs{k}},\tilde{m}_1,\tilde{m}_2;\bs{k},m_1,m_2} (\bs{p}) = S^{(\eta_1,\eta_2)*}_{\tilde{\bs{k}}+\bs{p},m_2,\tilde{m}_1;\bs{k}+\bs{p},\tilde{m}_2,m_1} (-\bs{p}-\bs{k}-\tilde{\bs{k}}),
\end{equation}
so that only one of them must be computed when performing numerical calculations.
It follows from Eqs.~\eqref{eq: rmatdef} and~\eqref{eq: neutralscattmat} that the charge $0$ scattering matrix at charge neutrality (filling $\nu = 0$) in Eq.~\eqref{eq: excitonscattmat} has exact zero modes at total momentum $\bs{p} = \bs{0}$:
\begin{equation} \label{eq: nonchiral_goldstone}
H \sum_{\bs{k},m} c^\dagger_{\bs{k},m,\eta,s_1} c_{\bs{k},m,\eta,s_2} \ket{\Psi_0} = 0.
\end{equation}
Here, $\eta$, $s_1$, and $s_2$ can be freely chosen as long as they do not annihilate $\ket{\Psi_0}$.

\subsection{$N=3$, $Q=1$}
We next move beyond \cite{TBG5, VafekKangPRB2021}  study trion excitations with charge $+1$. These were first studied in \cite{TBG6} by exact diagonalization in a restricted subspace. Here we further obtain their momentum structure. 
We begin by deriving the scattering matrix. At each total momentum $\bs{p}$, the charge $+1$ trions can be obtained by diagonalizing a $8N_{\mathcal{M}}^2 \times 8N_{\mathcal{M}}^2$ matrix. For the trion excitation $\mathcal{E} =c^\dagger_{\bs{k}_3,m_3,\eta_3,s_3} c^\dagger_{\bs{k}_2,m_2,\eta_2,s_2} c_{\bs{k}_2+\bs{k}_3-\bs{p},m_1,\eta_1,s_1}$, we obtain
\begin{equation} \label{eq: trionscattmat}
\begin{aligned}
H& c^\dagger_{\bs{k}_3,m_3,\eta_3,s_3} c^\dagger_{\bs{k}_2,m_2,\eta_2,s_2} c_{\bs{k}_2+\bs{k}_3-\bs{p},m_1,\eta_1,s_1} \ket{\Psi_0} =\\& \sum_{\tilde{\bs{k}}_3,\tilde{\bs{k}}_2,\tilde{m}_3,\tilde{m}_2,\tilde{m}_1} \bigg\{\delta_{\bs{k}_3,\tilde{\bs{k}}_3} \delta_{\bs{k}_2,\tilde{\bs{k}}_2} \bigg[R^{(\eta_3)}_{\tilde{m}_3,m_3} (\bs{k}_3) \delta_{\tilde{m}_2,m_2} \delta_{\tilde{m}_1,m_1} + R^{(\eta_2)}_{\tilde{m}_2,m_2} (\bs{k}_2) \delta_{\tilde{m}_3,m_3} \delta_{\tilde{m}_1,m_1} \\&+ R^{(\eta_1)}_{m_1,\tilde{m}_1} (\bs{k}_2+\bs{k}_3-\bs{p}) \delta_{\tilde{m}_2,m_2} \delta_{\tilde{m}_3,m_3}\bigg] + 2 \delta_{\tilde{m}_1,m_1} \delta_{\tilde{\bs{k}}_2+\tilde{\bs{k}}_3,\bs{k}_2+\bs{k}_3} T^{(\eta_3,\eta_2)}_{-\tilde{\bs{k}}_2,\tilde{m}_3,\tilde{m}_2;-\bs{k}_2,m_3,m_2} (\bs{k}_3+\bs{k}_2)
\\&-2 \delta_{\tilde{m}_2,m_2} \delta_{\tilde{\bs{k}}_2,\bs{k}_2} S^{(\eta_3,\eta_1)}_{\tilde{\bs{k}}_3+\bs{k}_2-\bs{p},\tilde{m}_3,\tilde{m}_1;\bs{k}_3+\bs{k}_2-\bs{p},m_3,m_1} (\bs{p}-\bs{k}_2)
-2 \delta_{\tilde{m}_3,m_3} \delta_{\tilde{\bs{k}}_3,\bs{k}_3} S^{(\eta_2,\eta_1)}_{\tilde{\bs{k}}_2+\bs{k}_3-\bs{p},\tilde{m}_2,\tilde{m}_1;\bs{k}_3+\bs{k}_2-\bs{p},m_2,m_1} (\bs{p}-\bs{k}_3) \bigg\}
\\&c^\dagger_{\tilde{\bs{k}}_3,\tilde{m}_3,\eta_3,s_3} c^\dagger_{\tilde{\bs{k}}_2,\tilde{m}_2,\eta_2,s_2} c_{\tilde{\bs{k}}_2+\tilde{\bs{k}}_3-\bs{p},m_1,\eta_1,s_1} \ket{\Psi_0} 
\\ \equiv& \sum_{\tilde{\bs{k}}_3,\tilde{\bs{k}}_2,\tilde{m}_3,\tilde{m}_2,\tilde{m}_1} W^{(\eta_3,\eta_2,\eta_1)}_{\tilde{\bs{k}}_3,\tilde{\bs{k}}_2,\tilde{m}_3,\tilde{m}_2,\tilde{m}_1; \bs{k}_3,\bs{k}_2,m_3,m_2,m_1} (\bs{p}) c^\dagger_{\tilde{\bs{k}}_3,\tilde{m}_3,\eta_3,s_3} c^\dagger_{\tilde{\bs{k}}_2,\tilde{m}_2,\eta_2,s_2} c_{\tilde{\bs{k}}_2+\tilde{\bs{k}}_3-\bs{p},m_1,\eta_1,s_1} \ket{\Psi_0},
\end{aligned}
\end{equation}
so that the Trion spectrum follows from diagonalizing the $8N_{\mathcal{M}}^2 \times 8N_{\mathcal{M}}^2$ matrix $W^{(\eta_3,\eta_2,\eta_1)}_{\tilde{\bs{k}}_3,\tilde{\bs{k}}_2,\tilde{m}_3,\tilde{m}_2,\tilde{m}_1; \bs{k}_3,\bs{k}_2,m_3,m_2,m_1} (\bs{p})$.
The corresponding scattering matrix for finite filling -- except for the third, diagonal term in Eq.~\eqref{eq: genericscattmat} -- is obtained from Eq.~\eqref{eq: trionscattmat} via the replacements:
\begin{equation}
R^{(\eta_3)}_{\tilde{m}_3,m_3} (\bs{k}_3) \rightarrow R^{(\eta_3,+)}_{\tilde{m}_3,m_3} (\bs{k}_3,E_{\bs{G}}), \quad 
R^{(\eta_2)}_{\tilde{m}_2,m_2} (\bs{k}_2) \rightarrow R^{(\eta_2,+)}_{\tilde{m}_2,m_2} (\bs{k}_2,E_{\bs{G}}), \quad
R^{(\eta_1)}_{m_1,\tilde{m}_1} (\bs{k}_2+\bs{k}_3-\bs{p}) \rightarrow R^{(\eta_1,-)}_{m_1,\tilde{m}_1} (\bs{k}_2+\bs{k}_3-\bs{p},E_{\bs{G}}),
\end{equation}
where $\tilde{R}^{(\eta,+)}_{n,m} (\bs{k},E_{\bs{G}})$ and $\tilde{R}^{(\eta,-)}_{n,m} (\bs{k},E_{\bs{G}})$ are defined in Eqs.~\eqref{eq: finitefillingRmatplus} and~\eqref{eq: finitefillingRmatminus}, respectively.

\section{Chiral limit}
To obtain the (first) chiral limit, we set $w_0 = 0$ for the BM model AA and BB hopping amplitude, where A and B are the two sublattices of the graphene honeycomb unit cell. In this limit, the form factors of Eq.~\eqref{eq: formfactors} become diagonal in the Chern basis~\cite{TBG1}, which is defined by the choice of creation operators
\begin{equation}
d^\dagger_{\bs{k},e_Y,\eta,s} = \frac{c^\dagger_{\bs{k},+,\eta,s}+\mathrm{i} e_Y c^\dagger_{\bs{k},-,\eta,s}}{\sqrt{2}}.
\end{equation}
Moreover, in the chiral limit, the form factors become independent of the valley index. Hence, we follow the notation of \cite{TBG5} and we denote them by $M^{(e_Y)}(\bs{k},\bs{q}+\bs{G})$. 

\begin{figure}[t]
\includegraphics[width=\textwidth]{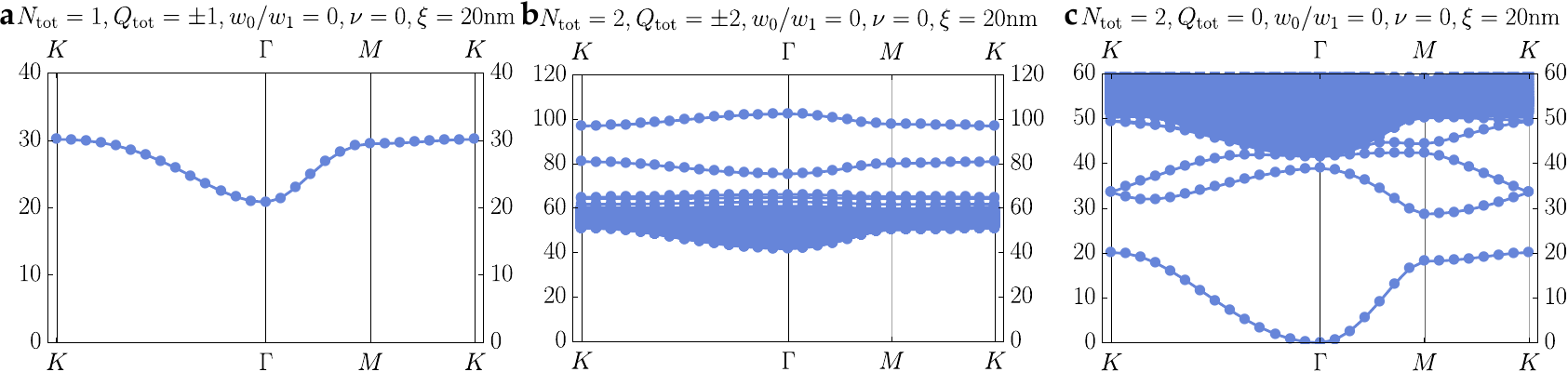}
\caption{Single particle and Goldstone spectra in the chiral limit. \textbf{a}~The spectrum of the matrix $R (\bs{k})$ of Eq.~\eqref{eq: rmatdef_chiral}. \textbf{b}~Spectrum of the chiral-limit $N=2$, $Q=2$ scattering matrix (Sec.~\ref{sec: n2q2scattmat}) for equal chiral sectors ($e_Y'=e_Y$). \textbf{c}~Spectrum of the chiral-limit $N=2$, $Q=0$ scattering matrix of Eq.~\eqref{eq: excitonscattmat_chiral} for equal chiral sectors ($e_Y'=e_Y$). The Goldstone mode is gapless at the $\Gamma$ point of the mBZ.}
\label{fig: goldstones}
\end{figure}

\subsection{Goldstone modes}
In the chiral limit, the $N=2$, $Q=0$ scattering equation at charge neutrality (Sec.~\ref{sec: excitons}) becomes
\begin{equation} \label{eq: excitonscattmat_chiral}
\begin{aligned}
H d^\dagger_{\bs{k}+\bs{p},e_Y,\eta_1,s_1} d_{\bs{k},e_Y',\eta_2,s_2} \ket{\Psi_0} = \sum_{\tilde{\bs{k}}} \bigg\{&\delta_{\bs{k},\tilde{\bs{k}}} \left[R (\bs{k}+\bs{p}) + R (\bs{k}) \right] -2S^{(e_Y,e_Y')}_{\tilde{\bs{k}};\bs{k}} (\bs{p}) \bigg\} d^\dagger_{\tilde{\bs{k}}+\bs{p},e_Y,\eta_1,s_1} d_{\tilde{\bs{k}},e_Y',\eta_2,s_2} \ket{\Psi_0}.
\end{aligned}
\end{equation}
Here, we have used the chiral limit one- and two-particle scattering matrices
\begin{equation} \label{eq: rmatdef_chiral}
\begin{aligned}
R (\bs{k}) &= \frac{1}{2\Omega} \sum_{\bs{q},\bs{G}} V(\bs{q}+\bs{G}) |M^{(e_Y)}(\bs{k},\bs{q}+\bs{G})|^2, \\
S^{(e_Y,e_Y')}_{\tilde{\bs{k}};\bs{k}} (\bs{p}) &= \frac{1}{2\Omega} \sum_{\bs{G}} V(\bs{k}-\tilde{\bs{k}}+\bs{G}) M^{(e_Y)*} (\tilde{\bs{k}}+\bs{p},\bs{k}-\tilde{\bs{k}}+\bs{G}) M^{(e_Y')} (\tilde{\bs{k}},\bs{k}-\tilde{\bs{k}}+\bs{G}).
\end{aligned}
\end{equation}
Note that the scalar $R (\bs{k})$ is independent of the Chern basis label $e_Y$~\cite{TBG3}.
At total momentum $\bs{p} = \bs{0}$ (the $\Gamma$ point of the mBZ), we can explicitly find the Goldstone modes [see Eq.~\eqref{eq: nonchiral_goldstone} for the expression away from the chiral limit]
\begin{equation}
\ket{G^{(e_Y,\eta_2,s_2,\eta_1,s_1)}_{\bs{p}=\bs{0}}} = \frac{1}{\sqrt{N_{\mathcal{M}}}} \sum_{\bs{k}} c^\dagger_{\bs{k},e_Y,\eta_2,s_2} c_{\bs{k}-\bs{p},e_Y,\eta_1,s_1} \ket{\Psi_0}.
\end{equation}
Away from $\Gamma$, the Goldstone modes have the general form
\begin{equation}
\ket{G^{(e_Y,\eta_2,s_2,\eta_1,s_1)}_{\bs{p}}} = \sum_{\bs{k}} G^{(e_Y)}_{\bs{p},\bs{k}} c^\dagger_{\bs{k},e_Y,\eta_2,s_2} c_{\bs{k}-\bs{p},e_Y,\eta_1,s_1} \ket{\Psi_0}.
\end{equation}
where we have introduced the Goldstone wave function $G^{(e_Y)}_{\bs{p},\bs{k}}$ that can be found numerically. For the parameters given in Sec.~\ref{sec: units_and_values}, we obtain the Goldstone dispersion in Fig.~\ref{fig: goldstones}\textbf{c}.

\begin{figure}[t]
\includegraphics[width=\textwidth]{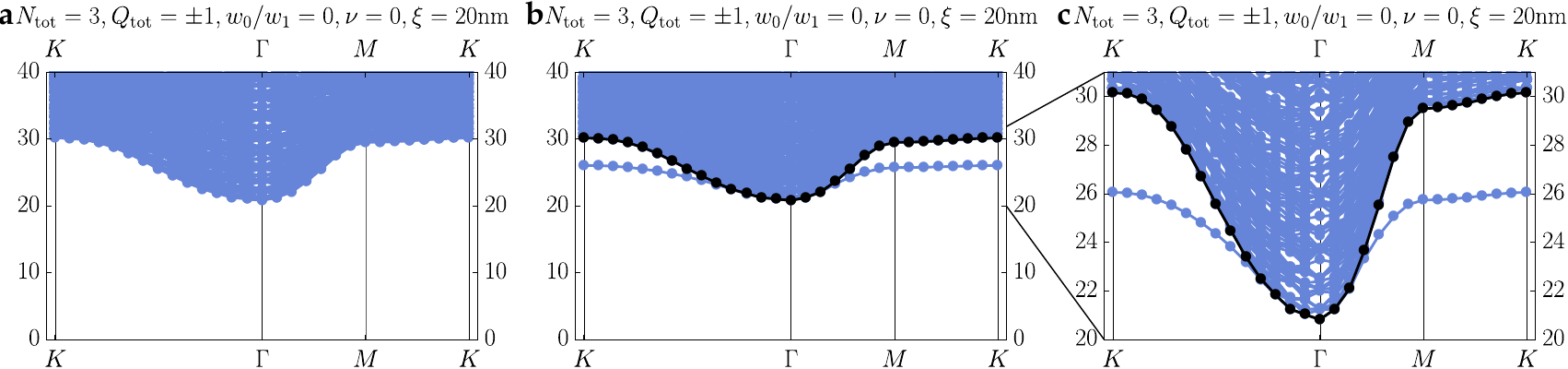}
\caption{Quasiparticle-Goldstone and trion spectra. \textbf{a}~Free quasiparticle-Goldstone continuum. \textbf{b}~Spectrum of the chiral-limit trion scattering matrix in Eq.~\eqref{eq: chiral_limit_trions} for equal chiral sectors ($e_Y''=e_Y'=e_Y$), with minimum of the quasiparticle-Goldstone continuum highlighted. \textbf{c}~Enlarged trion spectrum. A trion bound state develops away from the $\Gamma$ point of the mBZ.}
\label{fig: trions}
\end{figure}

\subsection{Full trion matrix}
In the chiral limit, the trion matrix simplifies to an $N_{\mathcal{M}}^2 \times N_{\mathcal{M}}^2$ matrix:
\begin{equation} \label{eq: trionscattmat_chiral}
\begin{aligned}
H& d^\dagger_{\bs{k}_3,e_Y'',\eta_3,s_3} d^\dagger_{\bs{k}_2,e_Y',\eta_2,s_2} d_{\bs{k}_2+\bs{k}_3-\bs{p},e_Y,\eta_1,s_1} \ket{\Psi_0} = \sum_{\tilde{\bs{k}}_3,\tilde{\bs{k}}_2} W^{(e_Y'',e_Y',e_Y)}_{\tilde{\bs{k}}_3,\tilde{\bs{k}}_2; \bs{k}_3,\bs{k}_2} (\bs{p}) d^\dagger_{\tilde{\bs{k}}_3,e_Y'',\eta_3,s_3} d^\dagger_{\tilde{\bs{k}}_2,e_Y',\eta_2,s_2} d_{\tilde{\bs{k}}_2+\tilde{\bs{k}}_3-\bs{p},e_Y,\eta_1,s_1} \ket{\Psi_0},
\end{aligned}
\end{equation}
where we have defined
\begin{equation} \label{eq: qmatdef_chiral}
\begin{aligned}
W^{(e_Y'',e_Y',e_Y)}_{\tilde{\bs{k}}_3,\tilde{\bs{k}}_2; \bs{k}_3,\bs{k}_2} (\bs{p}) =& \delta_{\bs{k}_3,\tilde{\bs{k}}_3} \delta_{\bs{k}_2,\tilde{\bs{k}}_2} \bigg[R (\bs{k}_3) + R (\bs{k}_2) + R (\bs{k}_2+\bs{k}_3-\bs{p}) \bigg] + 2 \delta_{\tilde{\bs{k}}_2+\tilde{\bs{k}}_3,\bs{k}_2+\bs{k}_3} T^{(e_Y'',e_Y')}_{-\tilde{\bs{k}}_2;-\bs{k}_2} (\bs{k}_3+\bs{k}_2)
\\&-2 \delta_{\tilde{\bs{k}}_2,\bs{k}_2} S^{(e_Y'',e_Y)}_{\tilde{\bs{k}}_3+\bs{k}_2-\bs{p};\bs{k}_3+\bs{k}_2-\bs{p}} (\bs{p}-\bs{k}_2)
-2 \delta_{\tilde{\bs{k}}_3,\bs{k}_3} S^{(e_Y',e_Y)}_{\tilde{\bs{k}}_2+\bs{k}_3-\bs{p};\bs{k}_3+\bs{k}_2-\bs{p}} (\bs{p}-\bs{k}_3).
\end{aligned}
\end{equation}
Here, we have used the chiral limit one- and two-particle scattering matrices
\begin{equation} \label{eq: tmatdef_chiral}
\begin{aligned}
R (\bs{k}) &= \frac{1}{2\Omega} \sum_{\bs{q},\bs{G}} V(\bs{q}+\bs{G}) |M^{(e_Y)}(\bs{k},\bs{q}+\bs{G})|^2, \\
T^{(e_Y,e_Y')}_{\tilde{\bs{k}};\bs{k}} (\bs{p}) &= \frac{1}{2\Omega} \sum_{\bs{G}} V(\bs{k}-\tilde{\bs{k}}+\bs{G}) M^{(e_Y)*} (\tilde{\bs{k}}+\bs{p},\bs{k}-\tilde{\bs{k}}+\bs{G}) M^{(e_Y')} (-\bs{k},\bs{k}-\tilde{\bs{k}}+\bs{G}),\\
S^{(e_Y,e_Y')}_{\tilde{\bs{k}};\bs{k}} (\bs{p}) &= \frac{1}{2\Omega} \sum_{\bs{G}} V(\bs{k}-\tilde{\bs{k}}+\bs{G}) M^{(e_Y)*} (\tilde{\bs{k}}+\bs{p},\bs{k}-\tilde{\bs{k}}+\bs{G}) M^{(e_Y')} (\tilde{\bs{k}},\bs{k}-\tilde{\bs{k}}+\bs{G}).
\end{aligned}
\end{equation}

\subsection{Goldstone-projected trion matrix}
From Fig.~\ref{fig: goldstones}, we see that the low energy part of the $N=3$, $Q=1$ Hilbert space is dominated by the combination of charge $+1$ ($N=1$, $Q=1$) and Goldstone ($N=2$, $Q=0$) excitations, hereafter called quasiparticle-Goldstone excitations. Conversely, the combination of charge $-1$ ($N=1$, $Q=-1$) and two-electron ($N=2$, $Q=2$) excitations is relatively costly as the charge excitation is gapped. Moreover, the lowest Goldstone mode is fully gapped from all higher Goldstone excitations, as is evident from Fig.~\ref{fig: goldstones}\textbf{c}. [Away from the chiral limit, however, this mode crosses an exciton mode ($N=2$, $Q=0$) that lies in a different chiral sector than the Goldstone modes ($e_Y \neq e_Y'$, not shown in figure).] Therefore, in the chiral limit, it is a good low-energy approximation to project the $N=3$, $Q=1$ scattering matrix into the variational quasiparticle-Goldstone basis
\begin{equation} \label{eq: varbasis}
\begin{aligned}
\ket{\Phi^{(e_Y',e_Y)}_{\bs{p};\bs{q},1}} =& \frac{1}{\sqrt{2}} \sum_{\bs{k}} G^{(e_Y)}_{\bs{p}-\bs{q},\bs{k}} \left(d^\dagger_{\bs{q},e_Y',\eta_3,s_3} d^\dagger_{\bs{k},e_Y,\eta_2,s_2} + d^\dagger_{\bs{k},e_Y,\eta_3,s_3} d^\dagger_{\bs{q},e_Y',\eta_2,s_2} \right) d_{\bs{q}+\bs{k}-\bs{p},e_Y,\eta_1,s_1} \ket{\Psi_0} \\
=& \frac{1}{\sqrt{2}} \left(d^\dagger_{\bs{q},e_Y',\eta_3,s_3} \ket{G^{(e_Y,\eta_2,s_2,\eta_1,s_1)}_{\bs{p}-\bs{q}}} 
-d^\dagger_{\bs{q},e_Y',\eta_2,s_2} \ket{G^{(e_Y,\eta_3,s_3,\eta_1,s_1)}_{\bs{p}-\bs{q}}} \right),
\\
\ket{\Phi^{(e_Y',e_Y)}_{\bs{p};\bs{q},2}} =& \frac{1}{\sqrt{2}} \sum_{\bs{k}} G^{(e_Y)}_{\bs{p}-\bs{q},\bs{k}} \left(d^\dagger_{\bs{q},e_Y',\eta_3,s_3} d^\dagger_{\bs{k},e_Y,\eta_2,s_2} - d^\dagger_{\bs{k},e_Y,\eta_3,s_3} d^\dagger_{\bs{q},e_Y',\eta_2,s_2} \right) d_{\bs{q}+\bs{k}-\bs{p},e_Y,\eta_1,s_1} \ket{\Psi_0}\\
=& \frac{1}{\sqrt{2}} \left(d^\dagger_{\bs{q},e_Y',\eta_3,s_3} \ket{G^{(e_Y,\eta_2,s_2,\eta_1,s_1)}_{\bs{p}-\bs{q}}} 
+d^\dagger_{\bs{q},e_Y',\eta_2,s_2} \ket{G^{(e_Y,\eta_3,s_3,\eta_1,s_1)}_{\bs{p}-\bs{q}}} \right).
\end{aligned}
\end{equation}
Here, we assume that the valley-spin flavors $(\eta_3,s_3) \neq (\eta_2,s_2) \neq (\eta_1,s_1)$ are chosen such that $(\eta_1,s_1)$ is occupied and $(\eta_3,s_3)$, $(\eta_2,s_2)$ are empty in $\ket{\Psi_0}$, and we have dropped the valley-spin labels on the left-hand side to minimize clutter. Importantly, this basis is not orthonormal: the overlap matrix reads ($\alpha,\beta=1,2$)
\begin{equation} \label{eq: varstate_overlapmat}
\begin{aligned}
\mathcal{O}^{(\bs{p},e_Y',e_Y)}_{\bs{q},\alpha;\bs{q}',\beta} = \braket{\Phi^{(e_Y',e_Y)}_{\bs{p};\bs{q},\alpha} | \Phi^{(e_Y',e_Y)}_{\bs{p};\bs{q}',\beta}} &= 
\begin{pmatrix} 
\delta_{\bs{q},\bs{q}'} + \delta_{e_Y,e_Y'} G^{(e_Y)*}_{\bs{p}-\bs{q},\bs{q}'} G^{(e_Y)}_{\bs{p}-\bs{q}',\bs{q}} & 0\\
0 & \delta_{\bs{q},\bs{q}'} - \delta_{e_Y,e_Y'} G^{(e_Y)*}_{\bs{p}-\bs{q},\bs{q}'} G^{(e_Y)}_{\bs{p}-\bs{q}',\bs{q}}
\end{pmatrix}_{\alpha, \beta} \\
&\equiv 
\left[\mathbb{1}_{N_{\mathcal{M}} \times N_{\mathcal{M}}} \otimes \sigma_0 + \delta_{e_Y,e_Y'} \Delta^{(\bs{p},e_Y)} \otimes \sigma_z\right]_{\bs{q},\alpha;\bs{q}',\beta},
\end{aligned}
\end{equation}
where $\Delta^{(\bs{p},e_Y)}$ is a $N_{\mathcal{M}} \times N_{\mathcal{M}}$ matrix with entries $$\Delta^{(\bs{p},e_Y)}_{\bs{q},\bs{q}'} = G^{(e_Y)*}_{\bs{p}-\bs{q},\bs{q}'} G^{(e_Y)}_{\bs{p}-\bs{q}',\bs{q}},$$ the $\sigma_i$, $i=0,x,y,z$, are the $2 \times 2$ Pauli matrices, and $\otimes$ denotes the Kronecker product.
When $e_Y'=e_Y$, the overlap matrix in Eq.~\eqref{eq: varstate_overlapmat} is not equal to the $2 N_{\mathcal{M}} \times 2 N_{\mathcal{M}}$ identity matrix, and this must be taken into account when calculating expectation values in the basis of Eq.~\eqref{eq: varbasis}. Specifically, the naive Hamiltonian matrix elements are given by
\begin{equation} \label{eq: hprojection_matrixel}
\begin{aligned}
&\mathcal{H}^{(\bs{p},e_Y',e_Y)}_{\bs{q},\alpha;\bs{q}',\beta} = \braket{\Phi^{(e_Y',e_Y)}_{\bs{p};\bs{q},\alpha} | H | \Phi^{(e_Y',e_Y)}_{\bs{p};\bs{q}',\beta}} =\\& 
\sum_{\bs{k},\bs{k}'} G^{(e_Y)*}_{\bs{p}-\bs{q},\bs{k}} G^{(e_Y)}_{\bs{p}-\bs{q}',\bs{k}'}
\begin{pmatrix} 
W^{(e_Y',e_Y,e_Y)}_{\bs{q},\bs{k}; \bs{q}',\bs{k}'} (\bs{p}) + \delta_{e_Y,e_Y'} W^{(e_Y,e_Y,e_Y)}_{\bs{k},\bs{q}; \bs{q}',\bs{k}'} (\bs{p}) & 0\\ 0 & W^{(e_Y',e_Y,e_Y)}_{\bs{q},\bs{k}; \bs{q}',\bs{k}'} (\bs{p}) - \delta_{e_Y,e_Y'} W^{(e_Y,e_Y,e_Y)}_{\bs{k},\bs{q}; \bs{q}',\bs{k}'} (\bs{p})
\end{pmatrix}_{\alpha, \beta} \\
&\equiv \left[\mathcal{A}^{(\bs{p},e_Y',e_Y)} \otimes \sigma_0 + \delta_{e_Y,e_Y'} \mathcal{B}^{(\bs{p},e_Y)} \otimes \sigma_z\right]_{\bs{q},\alpha;\bs{q}',\beta},
\end{aligned}
\end{equation}
where we have defined the two $N_{\mathcal{M}} \times N_{\mathcal{M}}$ matrices $\mathcal{A}^{(\bs{p},e_Y',e_Y)}$ and $\mathcal{B}^{(\bs{p},e_Y)}$.
However, the Hamiltonian acts as
\begin{equation}
H \ket{\Phi^{(e_Y',e_Y)}_{\bs{p};\bs{q},\alpha}} = \sum_{\bs{q}',\beta} \tilde{\mathcal{H}}^{(\bs{p},e_Y',e_Y)}_{\bs{q}',\beta;\bs{q},\alpha} \ket{\Phi^{(e_Y',e_Y)}_{\bs{p};\bs{q}',\beta}} + \cdots,
\end{equation}
where the dots $\cdots$ abbreviate states lying outside of the variational space of Eq.~\eqref{eq: varbasis} and we have defined the $2 N_{\mathcal{M}} \times 2 N_{\mathcal{M}}$ matrix
\begin{equation} \label{eq: chiral_limit_trions}
\tilde{\mathcal{H}}^{(\bs{p},e_Y',e_Y)} = \left[\mathcal{O}^{(\bs{p},e_Y',e_Y)}\right]^{-1} \mathcal{H}^{(\bs{p},e_Y',e_Y)}.
\end{equation}
This matrix is not Hermitian but its spectrum is real and coincides with the variational energies. Moreover, its right eigenvectors are the variational states in the non-orthogonal basis of Eq.~\eqref{eq: varbasis}.
We proceed by evaluating the individual terms of Eq.~\eqref{eq: hprojection_matrixel}:
\begin{equation} \label{eq: amatrixexpansion}
\begin{aligned}
\mathcal{A}^{(\bs{p},e_Y',e_Y)}_{\bs{q},\bs{q}'} = \sum_{\bs{k},\bs{k}'} &G^{(e_Y)*}_{\bs{p}-\bs{q},\bs{k}} G^{(e_Y)}_{\bs{p}-\bs{q}',\bs{k}'} W^{(e_Y',e_Y,e_Y)}_{\bs{q},\bs{k}; \bs{q}',\bs{k}'} (\bs{p}) = \\
&\delta_{\bs{q},\bs{q}'} \left[R(\bs{q}) + \epsilon(\bs{p}-\bs{q}) \right] 
+ 2 \sum_{\bs{k}} \left(G^{(e_Y)*}_{\bs{p}-\bs{q},\bs{q}'-\bs{k}} G^{(e_Y)}_{\bs{p}-\bs{q}',\bs{q}-\bs{k}} 
- G^{(e_Y)*}_{\bs{p}-\bs{q},\bs{p}-\bs{k}} G^{(e_Y)}_{\bs{p}-\bs{q}',\bs{p}-\bs{k}} \right) S^{(e_Y',e_Y)}_{\bs{q}-\bs{k};\bs{q}'-\bs{k}} (\bs{k}).
\end{aligned}
\end{equation}
Here, $\epsilon (\bs{p})$ is the energy of the Goldstone mode at total momentum $\bs{p}$ (plotted in Fig.~\ref{fig: goldstones}\textbf{c}). Furthermore, we have
\begin{equation}
\begin{aligned}
\mathcal{B}^{(\bs{p},e_Y)}_{\bs{q},\bs{q}'} = \sum_{\bs{k},\bs{k}'} &G^{(e_Y)*}_{\bs{p}-\bs{q},\bs{k}} G^{(e_Y)}_{\bs{p}-\bs{q}',\bs{k}'} W^{(e_Y,e_Y,e_Y)}_{\bs{k},\bs{q}; \bs{q}',\bs{k}'} (\bs{p}) = \\&G^{(e_Y)*}_{\bs{p}-\bs{q},\bs{q}'} G^{(e_Y)}_{\bs{p}-\bs{q}',\bs{q}} \left[R(\bs{q}') + R(\bs{q}) + R(\bs{q}'+\bs{q}-\bs{p}) \right] 
+2 \sum_{\bs{k}} \bigg[G^{(e_Y)*}_{\bs{p}-\bs{q},\bs{k}-\bs{q}} G^{(e_Y)}_{\bs{p}-\bs{q}',\bs{k}-\bs{q}'} T^{(e_Y,e_Y)}_{-\bs{q};\bs{q}'-\bs{k}} (\bs{k}) \\ &
- G^{(e_Y)*}_{\bs{p}-\bs{q},\bs{k}} G^{(e_Y)}_{\bs{p}-\bs{q}',\bs{q}} S^{(e_Y,e_Y)}_{\bs{k}+\bs{q}-\bs{p};\bs{q}'+\bs{q}-\bs{p}} (\bs{p}-\bs{q}) 
- G^{(e_Y)*}_{\bs{p}-\bs{q},\bs{q}'} G^{(e_Y)}_{\bs{p}-\bs{q}',\bs{k}} S^{(e_Y,e_Y)}_{\bs{q}+\bs{q}'-\bs{p};\bs{k}+\bs{q}'-\bs{p}} (\bs{p}-\bs{q}') 
\bigg].
\end{aligned}
\end{equation}

\begin{figure}[t]
\includegraphics[width=\textwidth]{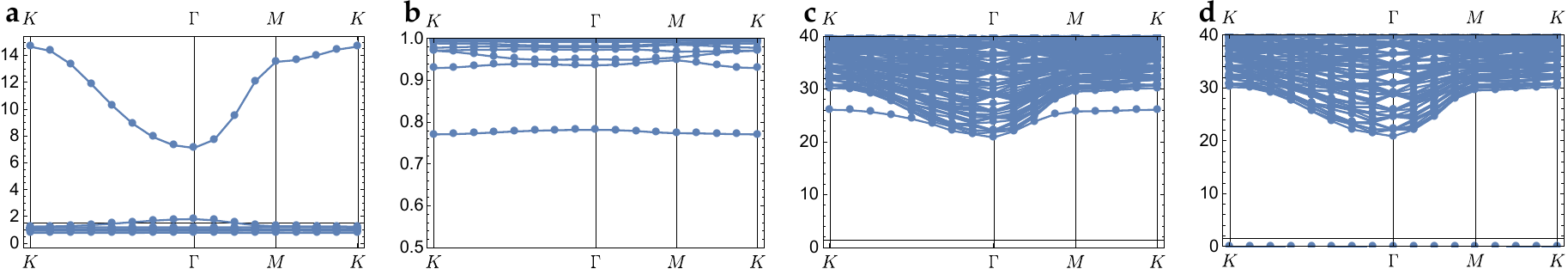}
\caption{Overlap matrix and trion bound states. \textbf{a}~Spectrum of the inverse overlap matrix $[\mathbb{1}_{N_{\mathcal{M}} \times N_{\mathcal{M}}} - \Delta^{(\bs{p},e_Y)}]^{-1}$ [Eq.~\eqref{eq: varstate_overlapmat}] for $\nu=0$ and double gate screening with $\xi=20$nm. \textbf{b}~Zoomed-in spectrum below eigenvalue $1$. \textbf{c}~Spectrum of the projected trion Hamiltonian $\tilde{\mathcal{H}}^{(\bs{p},e_Y,e_Y)}$ in Eq.~\eqref{eq: chiral_limit_trions}. \textbf{d}~Spectrum of the matrix $[\mathbb{1}-P(\bs{p})]\tilde{\mathcal{H}}^{(\bs{p},e_Y,e_Y)}[\mathbb{1}-P(\bs{p})]$.}
\label{fig: 7}
\end{figure}

\subsection{Analysis of the overlap matrix} \label{sec: overlapmat}
Numerically, we find that the trion bound states lie in the Chern basis sector $e_Y'=e_Y$ (see however Fig.~\ref{fig: 8} for spectra in the sector $e_Y'\neq e_Y$). This suggests that the bound states capitalize on the non-orthogonality of the variational states $\ket{\Phi^{(e_Y,e_Y)}_{\bs{p};\bs{q},1}}$ in this sector (in the sector $e_Y'\neq e_Y$, the variational states form an orthonormal basis). To prove this, we proceed to analyze the overlap matrix. The spectrum of the $\alpha=\beta=2$ block of the inverse overlap matrix $[\mathcal{O}^{(\bs{p},e_Y,e_Y)}]^{-1}$ is plotted in Fig.~\ref{fig: 7}\textbf{a},\textbf{b}. Recall that for an orthogonal basis, this spectrum would consist of a single band at eigenvalue $1$. In the present case, there are two notable deviations: (1) a dispersive gapped band at eigenvalue $\sim 8 \dots 14$, and (2) a flat gapped band at eigenvalue $\sim0.77$.

The presence of the band (1) can be understood by taking the limit where all Goldstone modes are constant $G^{(e_Y)}_{\bs{p},\bs{k}} \approx G^{(e_Y)}_{\bs{0},\bs{k}} = 1/\sqrt{N_{\mathcal{M}}}$. In this limit, the matrix $\Delta^{(\bs{p},e_Y)}_{\bs{q},\bs{q}'} = 1/N_{\mathcal{M}}$ has a single eigenstate at eigenvalue $1$, which is projected out by $[\mathcal{O}^{(\bs{p},e_Y,e_Y)}]^{-1}$ to infinite energy in the $\alpha =\beta=2$ sector of Eq.~\eqref{eq: varstate_overlapmat}. This is equivalent to the statement that in this limit the state $\sum_{\bs{q}} \ket{\Phi^{(e_Y,e_Y)}_{\bs{p};\bs{q},2}} = 0$ is unphysical and must be projected out to obtain an orthonormal basis. When restoring the $\bs{k}$-dependence of the Goldstone wavefunction $G^{(e_Y)}_{\bs{p},\bs{k}}$, this state becomes admissible and corresponds to the high overlap matrix band.

The band (2) directly implies the presence of a trion bound state. To see this, we compare the trion spectrum of $\tilde{\mathcal{H}}^{(\bs{p},e_Y,e_Y)}$ in Eq.~\eqref{eq: chiral_limit_trions} with the spectrum of the matrix 
\begin{equation}
[\mathbb{1}-P(\bs{p})]\tilde{\mathcal{H}}^{(\bs{p},e_Y,e_Y)}[\mathbb{1}-P(\bs{p})], 
\end{equation}
where $P(\bs{p})$ is the projector onto the low overlap band at eigenvalue $\sim0.77$. These spectra are plotted in Fig.~\ref{fig: 7}\textbf{c} and~\textbf{d}. We see that projecting out the flat overlap band amounts to projecting out the trion bound state, without much change to the rest of the spectrum. Hence, the presence of a trion bound state is tied to the lowest band of the inverse overlap matrix. In fact, this band has a $\mathcal{H}^{(\bs{p},e_Y',e_Y)}$ expectation value $\sim 35.8 \mathrm{meV}$, explaining the presence of a flat trion bound state at energy $\sim 0.77 \times 35.8 \mathrm{meV} = 27.6 \mathrm{meV}$ in the physical spectrum of $\tilde{\mathcal{H}}^{(\bs{p},e_Y',e_Y)}$. The actual bound state energy is slightly lower ($26.1 \mathrm{meV}$), because the low $[\mathcal{O}^{(\bs{p},e_Y',e_Y)}]^{-1}$ band is not an exact eigenstate of $\tilde{\mathcal{H}}^{(\bs{p},e_Y',e_Y)}$. The resulting flat bound state merges with the continuum when the quasiparticle energy dips below $26.1 \mathrm{meV}$ close to $\Gamma$.

\section{Numerical spectra}
\begin{figure}[t]
\includegraphics[width=\textwidth]{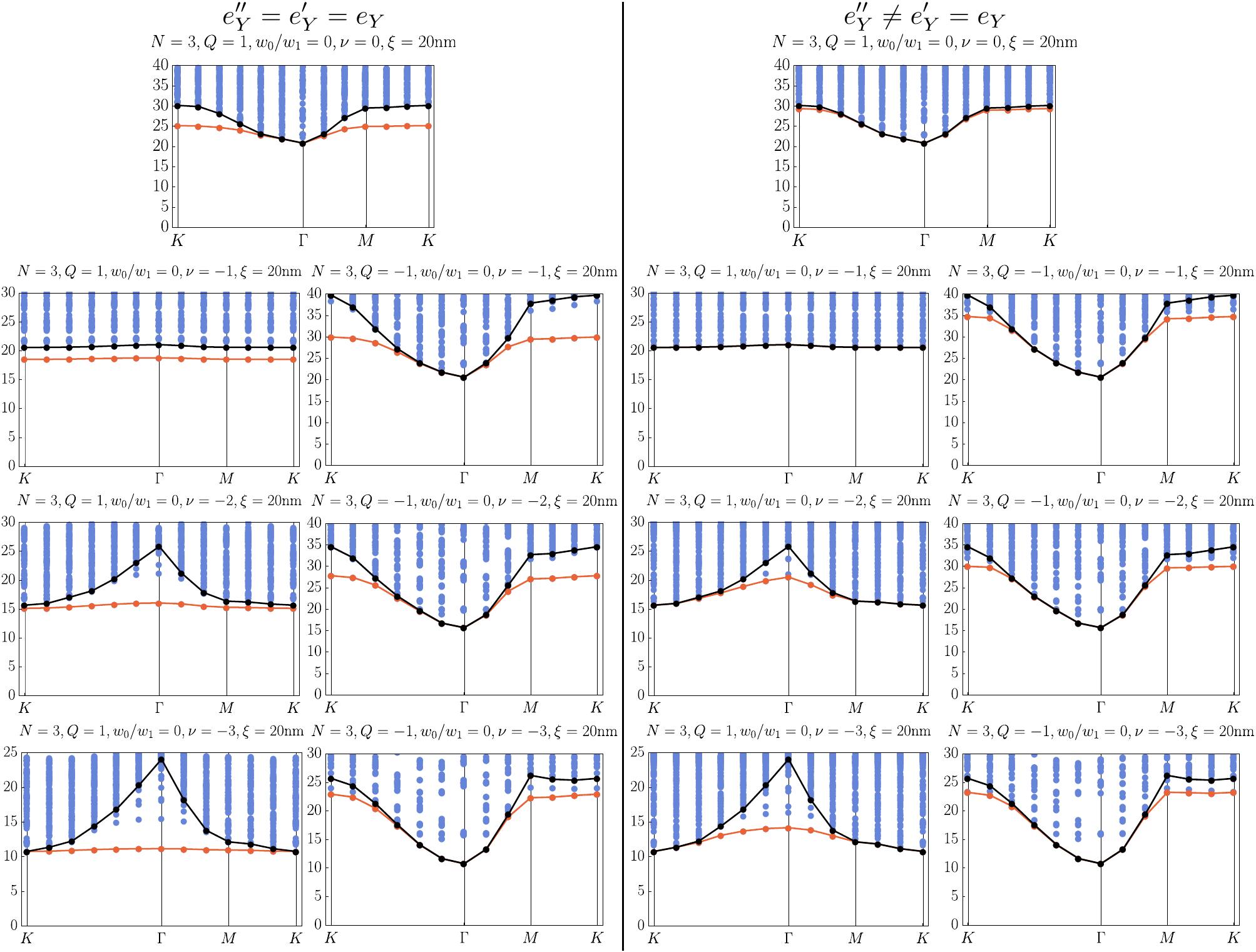}
\caption{Exact spectrum comparison between equal and opposite Chern sector trion modes [Eq.~\eqref{eq: qmatdef_chiral}]. All bound modes in the sectors with $e_Y''\neq e_Y' = e_Y$ have consistently higher energy than the corresponding modes in the sectors with $e_Y''= e_Y' = e_Y$.}
\label{fig: 8}
\end{figure}

Since most of our numerical trion spectra in the chiral limit are for the Chern sector $e_Y''= e_Y' = e_Y$ in Eq.~\eqref{eq: trionscattmat_chiral}, we provide spectra for $e_Y''\neq e_Y' = e_Y$ in Fig.~\ref{fig: 8} to show that there are no lower-energy bound states in those sectors.

\begin{figure}[t]
\includegraphics[width=\textwidth]{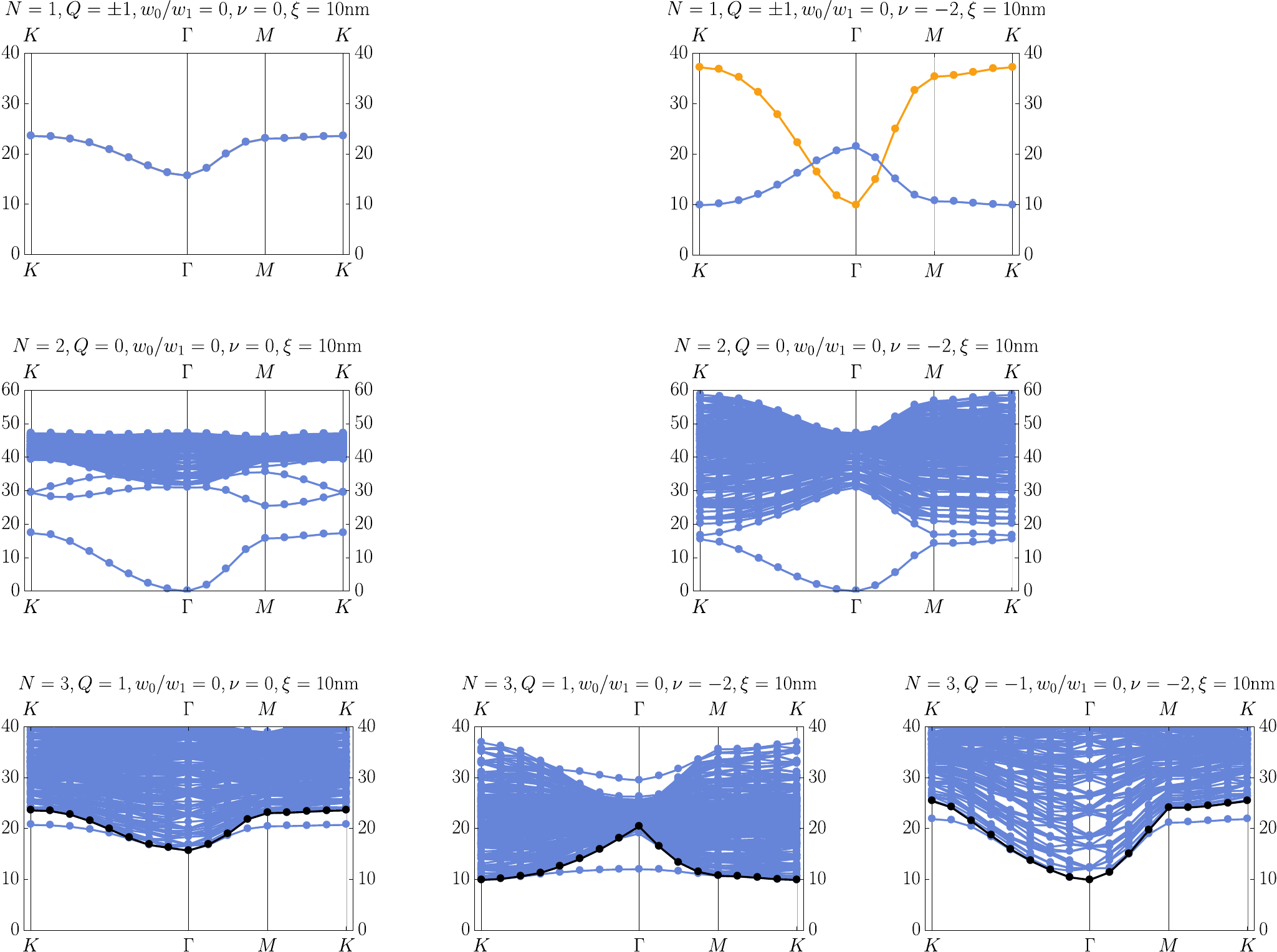}
\caption{Numerical spectra for a double gate-screened Coulomb potential at screening length $\xi=10$nm. For the trion spectra, we have used the quasiquasiparticle-Goldstone projection method [Eq.~\eqref{eq: chiral_limit_trions}].}
\label{fig: 9}
\end{figure}

\begin{figure}[t]
\includegraphics[width=\textwidth]{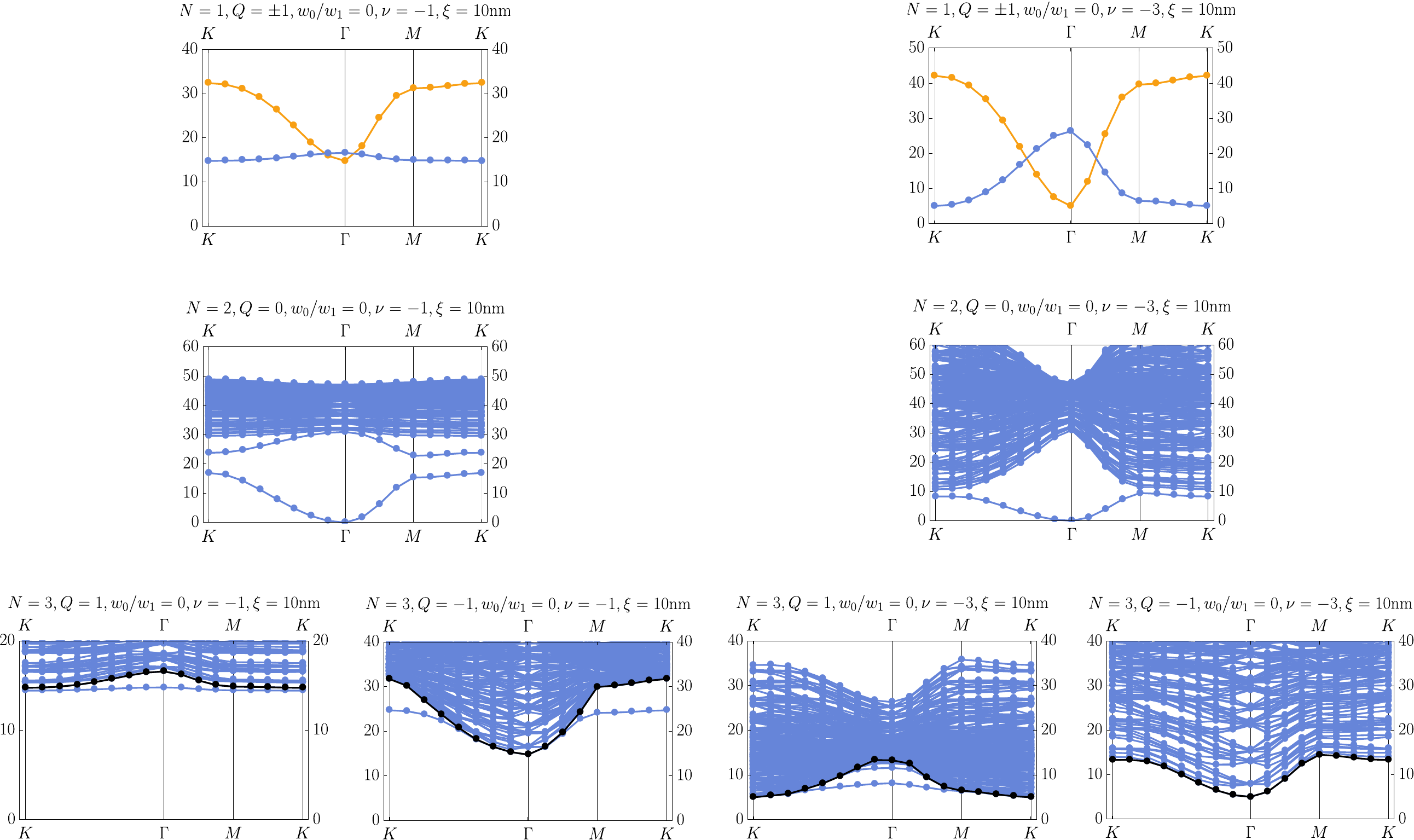}
\caption{Numerical spectra for a double gate-screened Coulomb potential at screening length $\xi=10$nm. For the trion spectra, we have used the quasiquasiparticle-Goldstone projection method [Eq.~\eqref{eq: chiral_limit_trions}].}
\label{fig: 10}
\end{figure}

\begin{figure}[t]
\includegraphics[width=\textwidth]{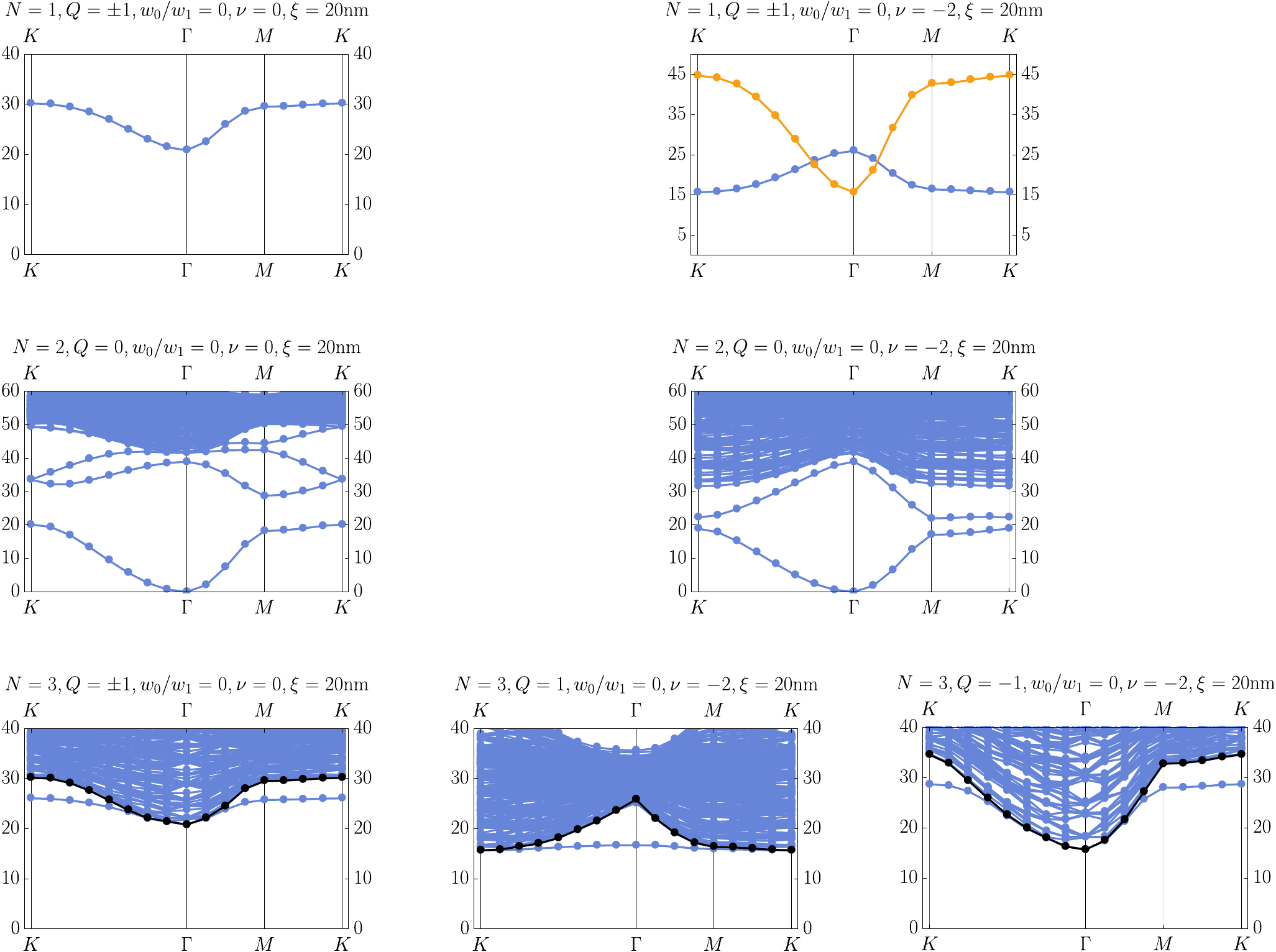}
\caption{Numerical spectra for a double gate-screened Coulomb potential at screening length $\xi=20$nm. For the trion spectra, we have used the quasiquasiparticle-Goldstone projection method [Eq.~\eqref{eq: chiral_limit_trions}].}
\label{fig: 11}
\end{figure}

\begin{figure}[t]
\includegraphics[width=\textwidth]{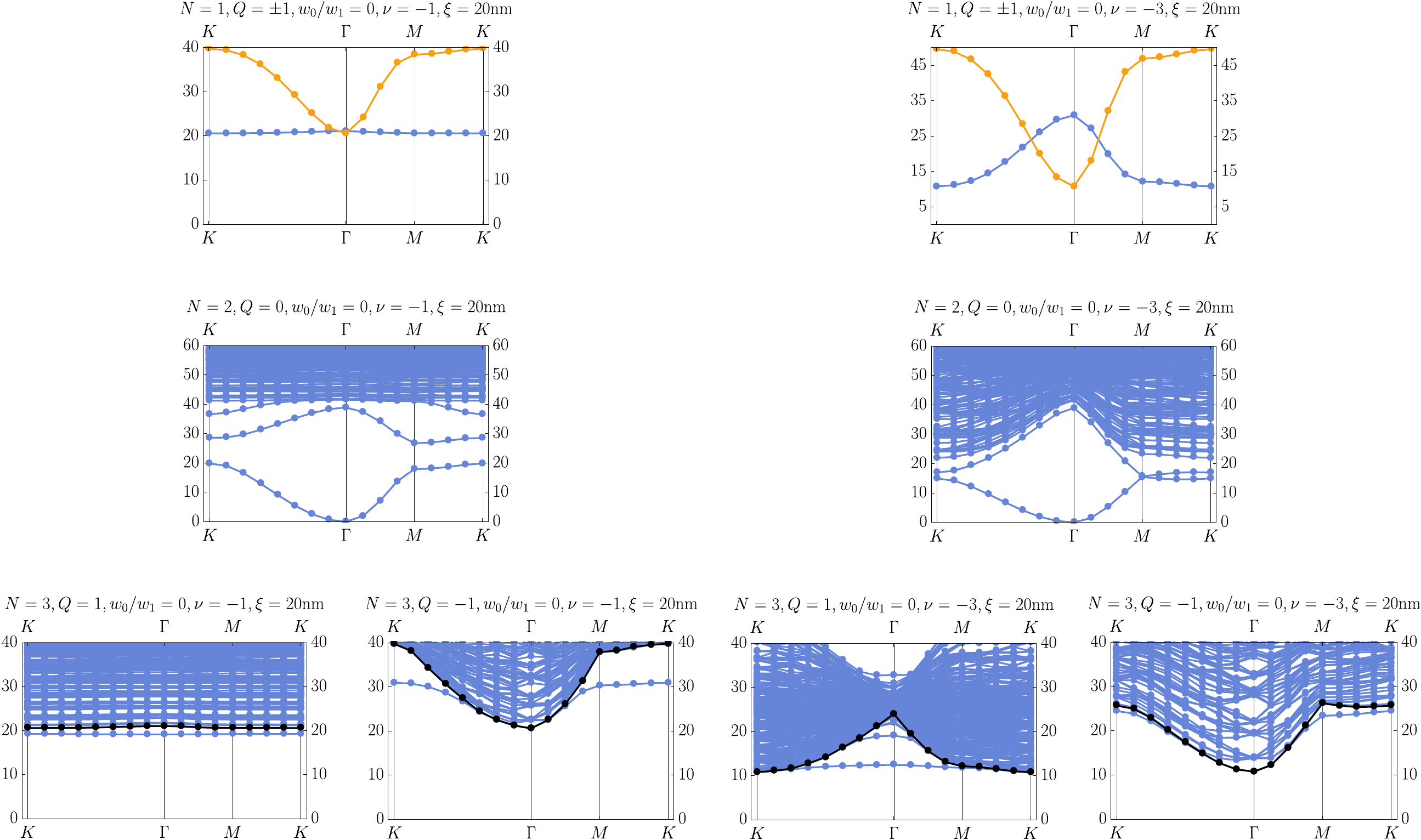}
\caption{Numerical spectra for a double gate-screened Coulomb potential at screening length $\xi=20$nm. For the trion spectra, we have used the quasiquasiparticle-Goldstone projection method [Eq.~\eqref{eq: chiral_limit_trions}].}
\label{fig: 12}
\end{figure}

Furthermore, we provide spectra for chiral limit $N=1,2,3$ excitations at all fillings and for a double gate-screened Coulomb potential with screening length $\xi=10$ and $\xi=20$ in Figs.~\ref{fig: 9}-\ref{fig: 12}.

\begin{figure}[t]
\includegraphics[width=\textwidth]{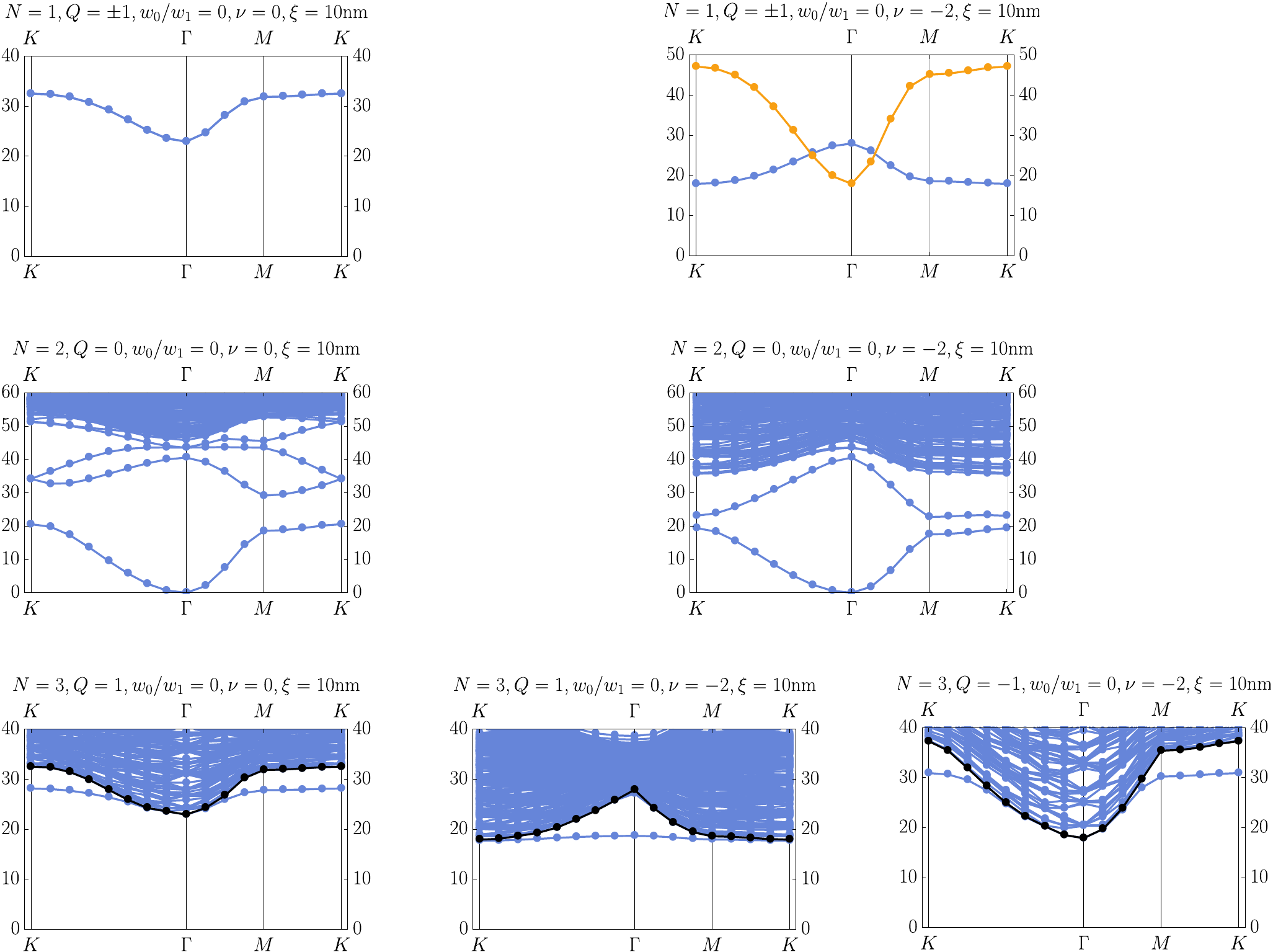}
\caption{Numerical spectra for a single gate-screened Coulomb potential at screening length $\xi=10$nm. For the trion spectra, we have used the quasiquasiparticle-Goldstone projection method [Eq.~\eqref{eq: chiral_limit_trions}].}
\label{fig: 13}
\end{figure}

\begin{figure}[t]
\includegraphics[width=\textwidth]{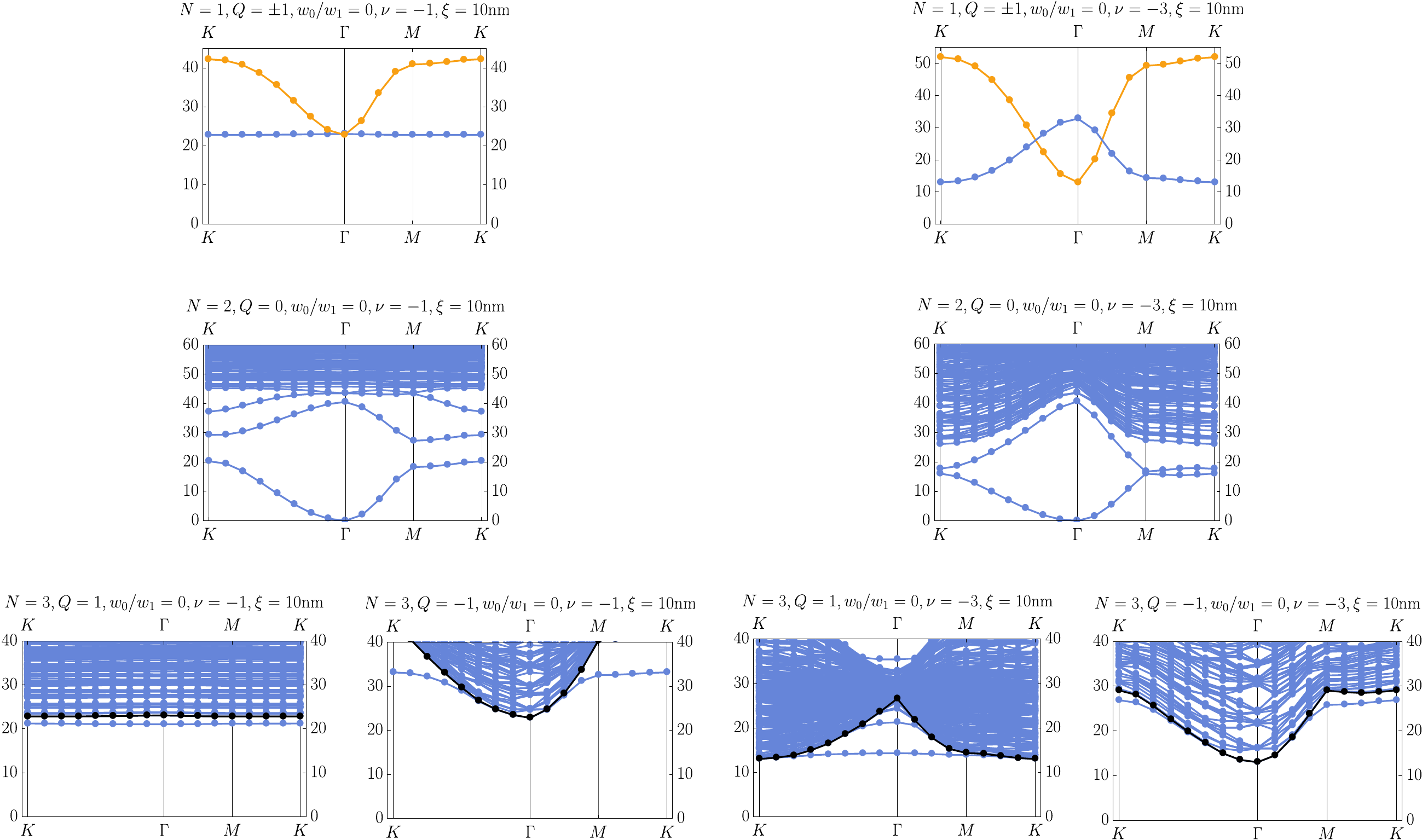}
\caption{Numerical spectra for a single gate-screened Coulomb potential at screening length $\xi=10$nm. For the trion spectra, we have used the quasiquasiparticle-Goldstone projection method [Eq.~\eqref{eq: chiral_limit_trions}].}
\label{fig: 14}
\end{figure}

Next to the double gate-screened Coulomb interaction defined in Eq.~\eqref{eq: coulombdoublegate}, we investigate the presence of trion bound states for the single gate-screened Coulomb interaction potential
\begin{equation}
    V(\bs{q}) 
    = 2\pi \xi^2 U_\xi \frac{1 - e^{-2\xi|\bs{q}|}}{\xi |\bs{q}|}.
\end{equation}
Importantly, here $\xi$ denotes the distance from the TBG sample to the single gate, whereas previously $\xi$ was the difference between the two gates, with the TBG sample located half-way in between. Hence, we compare the double gate $\xi=20$nm results with the single gate $\xi=10$nm results in Figs.~\ref{fig: 11}-\ref{fig: 14}.

\begin{figure}[t]
\includegraphics[width=\textwidth]{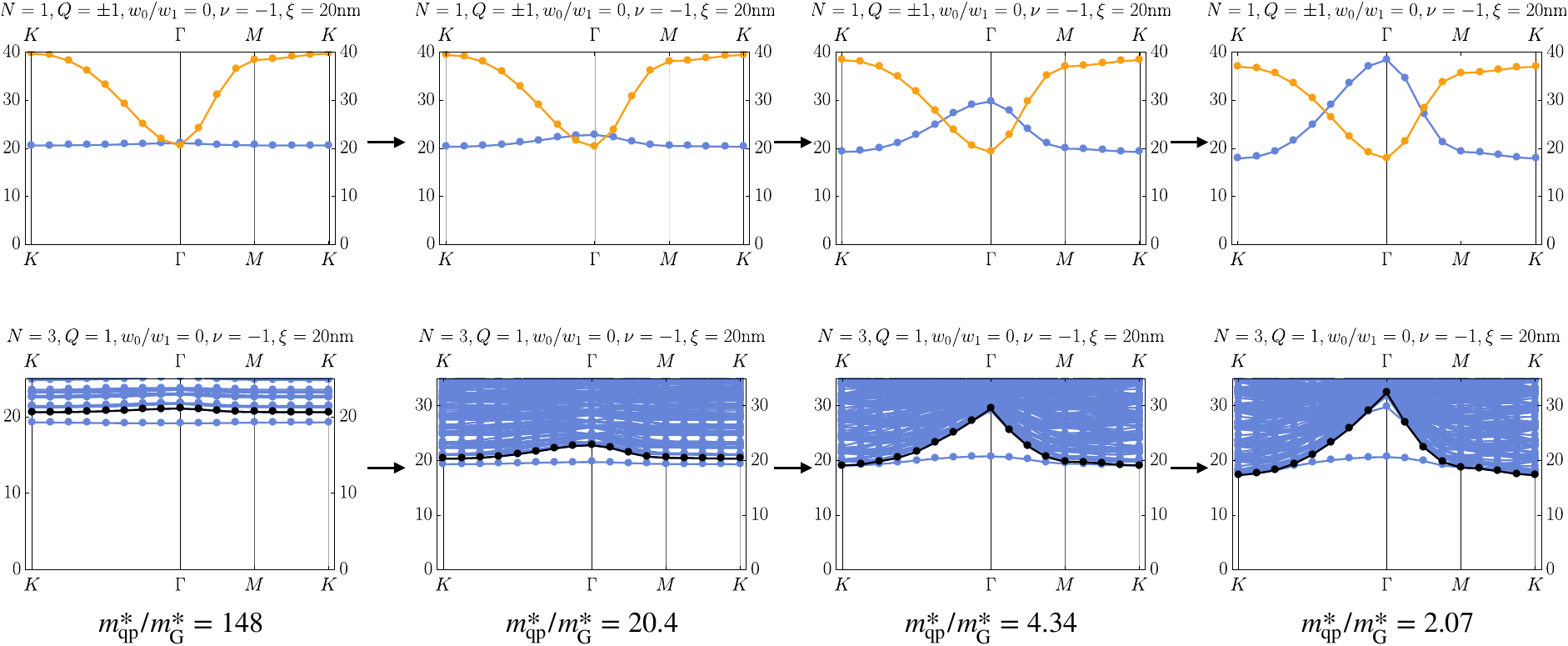}
\caption{Artificially tuning the quasiparticle dispersion at $\nu=-1, Q=1$ by hand to unbind the trion bound state. We assume double gate screening and use the quasiquasiparticle-Goldstone projection method [Eq.~\eqref{eq: chiral_limit_trions}]. To obtain this series of plots and Fig.~2\textbf{c}, we gradually add some dispersion from the $Q=+1$ quasiparticle at $\nu=-3$ to the $Q=+1$ quasiparticle at $\nu=-1$ (Fig.~\ref{fig: 12}, first row), thereby reducing the mass ratio $m^*_{\mathrm{qp}}/m^*_{\mathrm{G}}$.}
\label{fig: 15}
\end{figure}

In Fig.~\ref{fig: 15}, we show how the $\nu=-1$,$Q=1$ trion bound state becomes unbound at $\Gamma$ when the $Q=1$ quasiparticle effective mass is gradually lowered.

\begin{figure}[t]
\includegraphics[width=\textwidth]{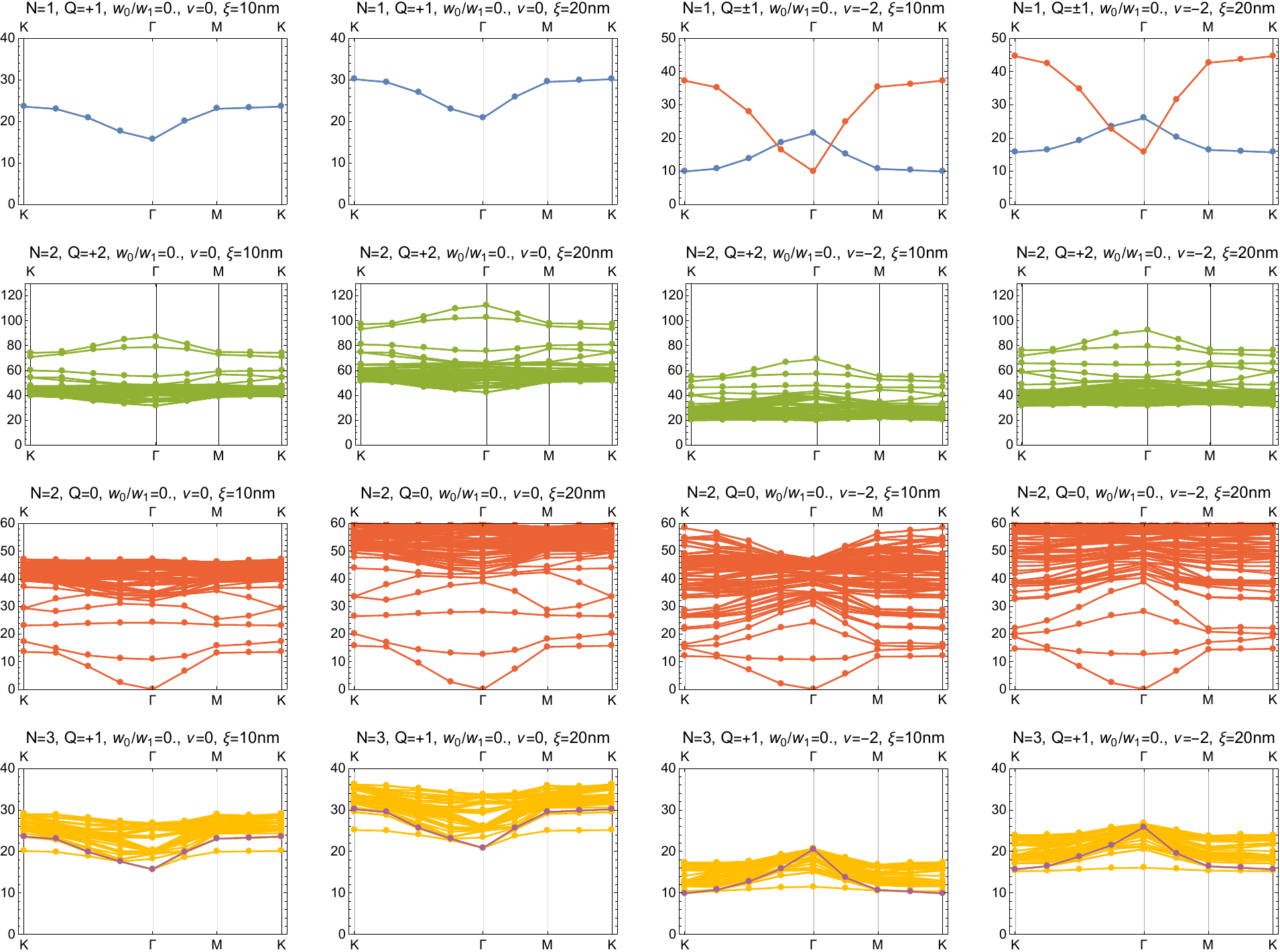}
\caption{Exact spectra in the chiral limit ($w_0/w_1=0$) for double gate screening with $\xi=10$nm and $\xi=20$nm, obtained without further resolving different Chern basis ($e_Y$) sectors [Eq.~\eqref{eq: trionscattmat}]. For the $N=3,Q=1$ trion plots, the minimum of the quasiquasiparticle-Goldstone continuum is highlighted (purple).}
\label{fig: 16}
\end{figure}

\begin{figure}[t]
\includegraphics[width=\textwidth]{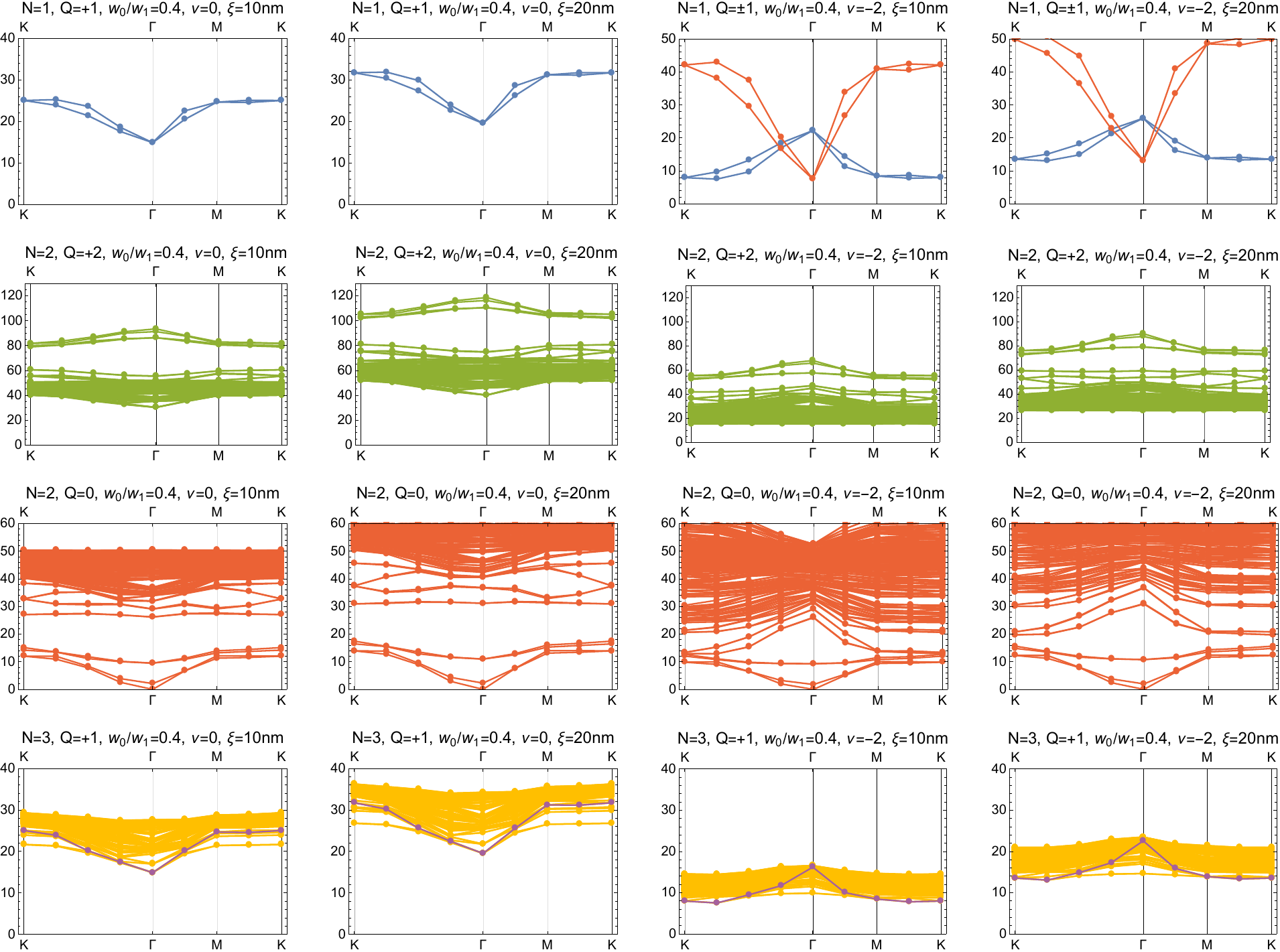}
\caption{Exact spectra [Eq.~\eqref{eq: trionscattmat}] for $w_0/w_1=0.4$ for double gate screening with $\xi=10$nm and $\xi=20$nm. For the $N=3,Q=1$ trion plots, the minimum of the quasiquasiparticle-Goldstone continuum is highlighted (purple).}
\label{fig: 17}
\end{figure}

\begin{figure}[t]
\includegraphics[width=\textwidth]{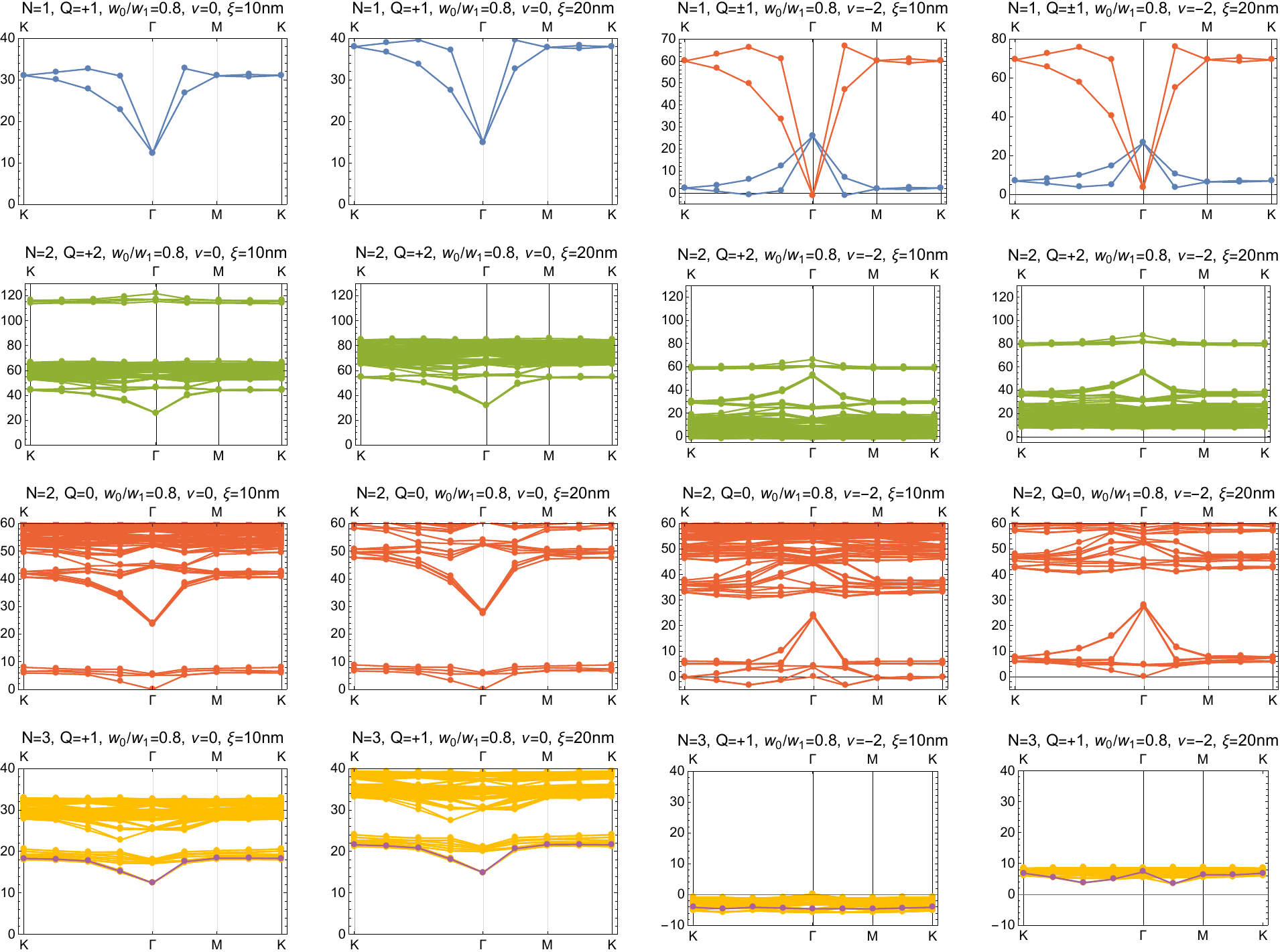}
\caption{Exact spectra [Eq.~\eqref{eq: trionscattmat}] for $w_0/w_1=0.8$ for double gate screening with $\xi=10$nm and $\xi=20$nm. For the $N=3,Q=1$ trion plots, the minimum of the quasiquasiparticle-Goldstone continuum is highlighted (purple). For $\xi=10$nm at $\nu=-2$, some excitations have negative energy and the ground state is unstable.}
\label{fig: 18}
\end{figure}

In Figs.~\ref{fig: 16}-\ref{fig: 18} we present $N=1,Q=\pm1$, $N=2,Q=0,2$, and $N=3,Q=1$ spectra at fillings $\nu=0,-2$ and tunneling ratios $w_0/w_1 = 0, 0.4, 0.8$ for double gate screening with $\xi=10$nm and $\xi=20$nm.

We find that the lowest trion bound states exclusively appear in the chiral sector $e_Y''= e_Y' = e_Y$ [Eq.~\eqref{eq: qmatdef_chiral}], as expected from our overlap matrix analysis in Sec.~\ref{sec: overlapmat}. Quite generally, we also find that larger screening lengths $\xi$ and single gate screening lead to larger trion binding energies than shorter $\xi$ and double gate screening. Moreover, all bound states vanish into the continuum as the tunneling ratio $w_0/w_1$ is increased to the realistic value $w_0/w_1 = 0.8$.

\begin{figure}[t]
\includegraphics[width=\textwidth]{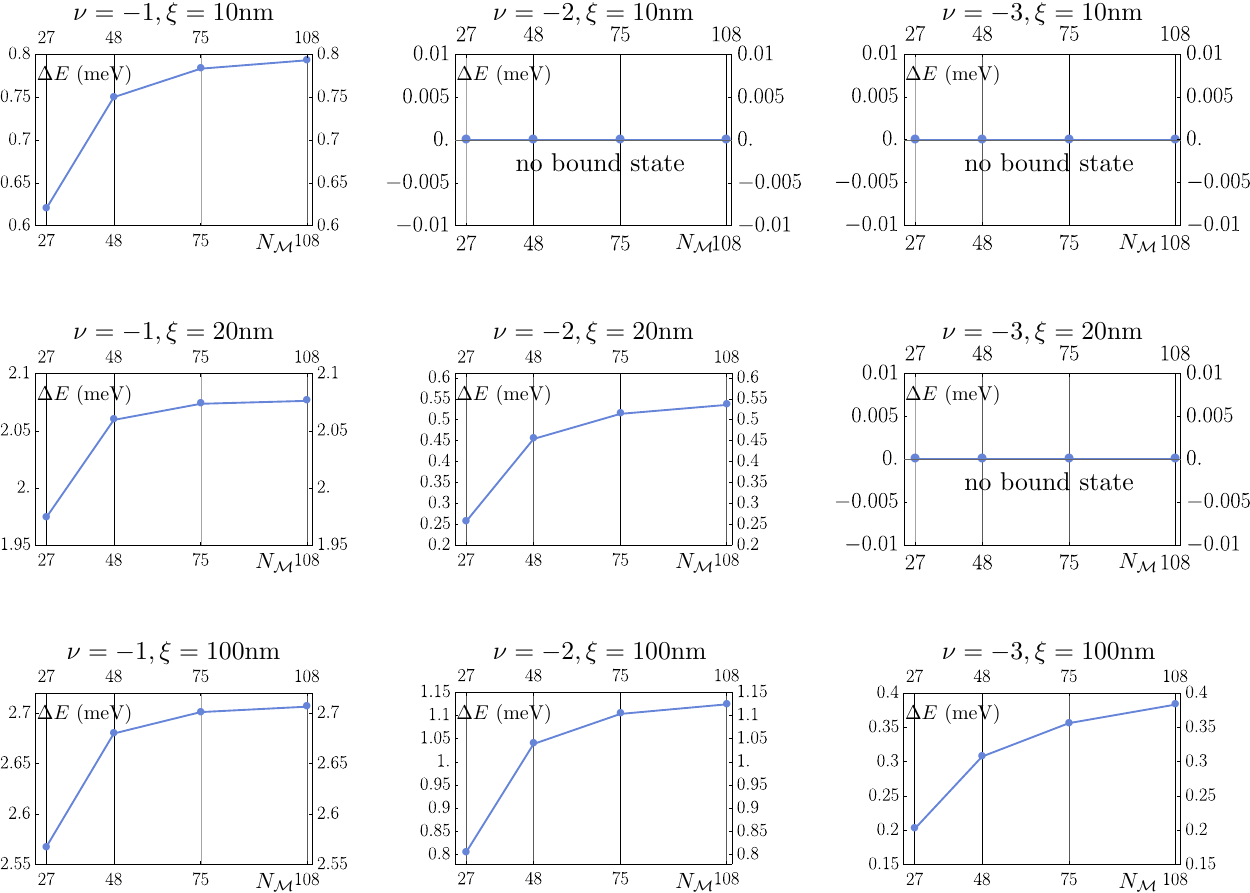}
\caption{Finite-size analysis showing the scaling of the trion binding energy $\Delta E$ with mBZ size $N_{\mathcal{M}}$ in the chiral limit ($w_0/w_1 = 0$). We use a mBZ with $N_{\mathcal{M}} = 3n^2 = 27,48,75,108$ sites that contains the $K$ point and preserves $\mathcal{C}_3$ rotational symmetry. For all screening lengths, the binding energy difference between the largest and second-largest mBZ is smaller than $0.05$meV.}
\label{fig: finitesize}
\end{figure}

Finally, in Fig.~\ref{fig: finitesize} we perform a finite-size analysis to show that our trion binding energies obtained from exact diagonalization in the chiral limit have converged to an error of within $0.05$meV for accessible mBZ sizes.

\clearpage
\bibliography{references}

\end{document}